\pdfoutput=1
\documentclass{aa}  
\usepackage{natbib}

\usepackage{graphicx}
\usepackage{txfonts}
\usepackage{lipsum}
\usepackage{subcaption}         
\usepackage{lscape}             
\usepackage{placeins}           

\usepackage{amsmath, amssymb}
\usepackage{mathrsfs}
\usepackage{upgreek}

\usepackage[colorlinks=true, linkcolor=black, citecolor=blue, urlcolor=blue]{hyperref}

\begin{document}

    \title{Tipping the scales in Blue Jay: Dust energy balance \\overpredicts IR luminosity in galaxies at $z \sim 2$}



   \author{Benjamin P. Collins\inst{1,2}\fnmsep\thanks{e-mail: \href{mailto:bcollins@das.uchile.cl}{bcollins@das.uchile.cl}}
        \and Sirio Belli\inst{1}
        \and Letizia Bugiani \inst{1,3}
        \and Amir H. Khoram \inst{1,3}
        \and Minjung Park \inst{4,5}
        \and Vanessa M. Brown \inst{6}
        \and Joel Leja \inst{7}
        \and Caterina Liboni \inst{1, 3}
        \and Gabriel Maheson \inst{1}
        \and Matteo Sapori \inst{1, 3}
        \and Charlotte Simmonds \inst{2}
        \and Anita Zanella \inst{3} 
        }

    \titlerunning{Dust energy balance overpredicts IR luminosity in galaxies at $z \sim 2$}
    \authorrunning{Collins et al.}

   \institute{Dipartimento di Fisica e Astronomia, Università di             Bologna, Bologna, Italy
            \and Departamento de Astronomía, Universidad de Chile, Camino El Observatorio 1515, Las Condes, Santiago, Chile
            \and INAF – OAS, Osservatorio di Astrofisica e Scienza dello Spazio di Bologna, via Gobetti 93/3, I-40129 Bologna, Italy
            \and Kavli Institute for Cosmology, University of Cambridge, Madingley Road, Cambridge CB3 0HA, UK
            \and Cavendish Laboratory, University of Cambridge, 19 JJ Thomson Avenue, Cambridge CB3 0HE, UK
            \and Department for Astrophysical and Planetary Science, University of Colorado, Boulder, CO, USA
            \and Department of Astronomy \& Astrophysics, The Pennsylvania State University, University Park, 16802, PA, USA.
            }

   \date{Submitted to A\&A}

 
  \abstract{

    The spectral energy distribution (SED) of a galaxy encodes vital information about its chemical composition, stellar populations, and dust content. To model the infrared emission, many modern SED-fitting codes rely on dust energy balance, which assumes that all energy absorbed by dust in the UV-optical is equal to the energy emitted by dust in the infrared regime. However, few studies have been carried out to test the validity of this assumption at high redshift, where direct infrared observations are sparse. In this work, we perform aperture photometry in 111 galaxies from the Blue Jay survey \mbox{($1.7 < z < 3.5)$} using public JWST/MIRI observations from the PRIMER, COSMOS-Web and COSMOS-3D surveys. Focusing on a subsample of 89 star-forming galaxies, we reconstruct best-fit SED model spectra from \textsc{Prospector} using only HST and NIRCam photometry and compare the predicted mid-infrared fluxes to the measured MIRI photometry. We find that mid-infrared fluxes are systematically overpredicted by about 10\% at 7.7$\upmu$m up to a factor of~$\sim3$ at 21$\upmu$m. The discrepancy increases with inferred dust attenuation, indicating that dust energy balance suffers from line-of-sight effects in systems with complex dust geometries. We find that adding NIRSpec spectroscopy to the fits breaks parameter degeneracies and strongly reduces the overprediction. When fitting the measured MIRI photometry together with the HST and NIRCam data, the best-fit model spectra of dusty galaxies ($A_V\gtrsim1$) result in significantly lower infrared luminosities by -0.6 dex compared to best-fit spectra obtained without MIRI, consistent with steeper, SMC-like attenuation curves or complex star--dust geometries. Our study highlights the limitations of dust energy balance at high redshift and underlines the importance of direct observations for accurately recovering the ISM and dust properties of galaxies at cosmic noon.

}



   \maketitle
   \nolinenumbers

\section{Introduction}

In recent decades, a large number of surveys have found that galaxies evolve rapidly at $1 < z < 3$, with about half of all stellar mass formed by $z\sim1$ \citep[e.g.][]{dickinson2003evolution, fontana2003assembly, rudnick2003luminosity, bundy2005mass, drory2005smf}. This crucial phase, commonly referred to as cosmic noon, is marked by the peak of the cosmic star-formation rate density and dust-obscured star formation \citep{casey2018alma, casey2021obscuration}, as well as a vast diversity of galaxy morphologies and colours \citep[e.g.][]{bundy2005mass, shapley2011physical, madau2014cosmic}. 

Our primary source of information about the physical properties of galaxies at low and high redshift is their integrated spectral energy distribution (SED), which is predominantly shaped by the stellar emission and its reprocessing by dust.  
Dust grains absorb and scatter stellar radiation in the UV-optical regime and re-emit this energy through a modified blackbody in the infrared \citep{draine2003grains, dacunha2010newinsights, orellana2017molecular}, especially in the most massive systems \citep{whitaker2017obscured}. Furthermore, the reddening of the UV continuum slope of galaxy spectra suffers from a strong degeneracy between dust attenuation, stellar age and metallicity \citep{papovich2001lyman, gallazzi2005degeneracy}.
To extract the wealth of information encoded in the SEDs of galaxies and connect observations to inferred physical properties, sophisticated models are needed.
Since full multi-wavelength sampling of galaxy SEDs is rare, SED models are constantly facing the task of extrapolating galaxy properties based on the limited information available (see reviews by \citealp[]{walcher2011fitting}, \citealp[]{conroy2013modeling}, and \citealp[]{salim2020dustreview}).

Early efforts to model the reprocessing of stellar light by dust relied on solving the radiative transfer equation for idealised distributions of stars and dust \citep{rowan1980radiative, efstathiou1990radiative, popescu2000radiative, gordon2001radiative}. More sophisticated approaches incorporated the evolutionary histories of stellar populations alongside dust, providing more realistic predictions of galaxy SEDs \citep{silva1998dust}. While these models are highly effective at interpreting the detailed emission of individual galaxies, their computational complexity makes them poorly suited for deriving statistical constraints from large galaxy samples \citep{dacunha2008simple}. Instead, many recent SED libraries and fitting frameworks such as \textsc{Cigale} \citep{noll2009cigale, boquien2019cigale}, \textsc{Magphys} \citep{dacunha2008simple, daCunha2011magphys}, and \textsc{Prospector} \citep{johnson2021prospector} rely on dust energy balance, which assumes that the energy absorbed in the rest-frame UV-optical regime is equal to the energy re-emitted by dust in the infrared. Hence, in principle, this approach allows constraining the infrared luminosity ($L_{\mathrm{IR}}$) by fitting the rest-frame UV-NIR regime alone. This is especially important as a precise estimate of $L_{\mathrm{IR}}$ is necessary to accurately infer the star formation history (SFH) and dust content of a galaxy \citep{leja2017deriving}. However, while providing an efficient framework for modelling galaxy SEDs, this approach comes with some key limitations. Crucially, measuring the starlight attenuation along a specific line of sight does not reliably constrain the global, isotropic distribution of dust emission, particularly in structurally asymmetric or highly inclined systems \citep{chevallard2013inclination}. Furthermore, galaxies exhibit a variety of dust geometries and grain size distributions which are difficult to generalise for large samples (\citealp[e.g.][]{zhang2023geometry}).

The unprecedented resolution and sensitivity of the James Webb Space Telescope (JWST; \citealp[]{gardner2023jwst}) in the near- and mid-infrared have revolutionised our ability to model high-redshift galaxy SEDs by directly probing their rest-frame UV-optical emission. Specifically, the Mid-Infrared Instrument (MIRI; \citealp[]{rieke2015miri}) allows us to probe the dust emission at cosmic noon in moderately star-forming galaxies, which was previously only possible in the most extreme systems.


In this paper, we assess the robustness of the dust energy balance assumption by comparing observed mid-infrared photometry from JWST/MIRI directly against predictions derived from optical and near-infrared observations. By fitting the rest-frame UV-to-near-infrared SED using only HST and JWST/NIRCam data, the mid-infrared dust emission is purely extrapolated from the inferred dust attenuation. This setup provides a powerful test of the energy balance assumption, allowing us to evaluate its predictive accuracy and identify potential limitations in current models describing the interplay between stellar populations, gas, and dust. 
We describe the JWST/MIRI observations and sample selection in Section \ref{sc:data}. The photometric workflow and the Blue Jay MIRI catalogue are presented in Section \ref{sc:photometry}. In Section \ref{sc:investigating_dust_energy_balance}, we test the performance of dust energy balance in predicting the measured MIRI fluxes by fitting different (spectro-)photometric datasets with \textsc{Prospector}. We summarise the results and conclude in Section~\ref{sc:discussion}. Throughout this paper we adopt a standard flat $\Lambda$CDM
cosmology with $\mathrm{H_0} = 67.7\,\mathrm{kms}^{-1}$, $\Omega_m = 0.31$ and  $\Omega_\Lambda= 0.69$ \citep{planck2020cosmo}.
\\

\section{Data}  \label{sc:data}
\subsection{The Blue Jay Survey}
The galaxies studied in this work were selected from the Blue Jay survey (GO 1810; \citealt{belli2026bluejaysurveydeep}), a medium-sized JWST Cycle 1 programme designed to study the interstellar medium and outflow-driven star formation quenching in galaxies at cosmic noon (\citealp[e.g.][]{belli2024agn, liboni2026neutral, khoram2026environments}). The survey obtained deep multi-object spectroscopy for 153 galaxies with two NIRSpec micro-shutter assembly (MSA) pointings in the Cosmic Evolution Survey (COSMOS) field with central coordinates (RA, Dec) = (150.125, 2.333) deg. Three medium-resolution NIRSpec gratings (G140M, G235M, G395M) were used to obtain spectra in the 1-5$\upmu$m range.

Elliptical aperture photometry was performed on multi-band NIRCam imaging from the PRIMER (Public Release IMaging for Extragalactic Research) survey (GO 1837; \citealt{dunlop2021primer}), which includes seven broad-band filters (F090W, F115W, F150W, F200W, F277W, F356W, and F444W) and one medium-band filter (F410M). In addition, multi-band HST imaging with ACS (F606W and F814W) and WFC3 (F125W, F140W, F160W) has been obtained in the COSMOS field as part of the CANDELS survey \citep{grogin2011,koekemoer2011}. The resulting photometric catalogue for the Blue Jay targets thus includes HST and JWST photometry in 13 bands, covering a wavelength range from 6000 Å to 4.4 $\upmu$m, which corresponds to a complete rest-frame coverage from the UV to the near-IR.

\begin{figure}[t]
    \centering
    \includegraphics[width=\hsize]{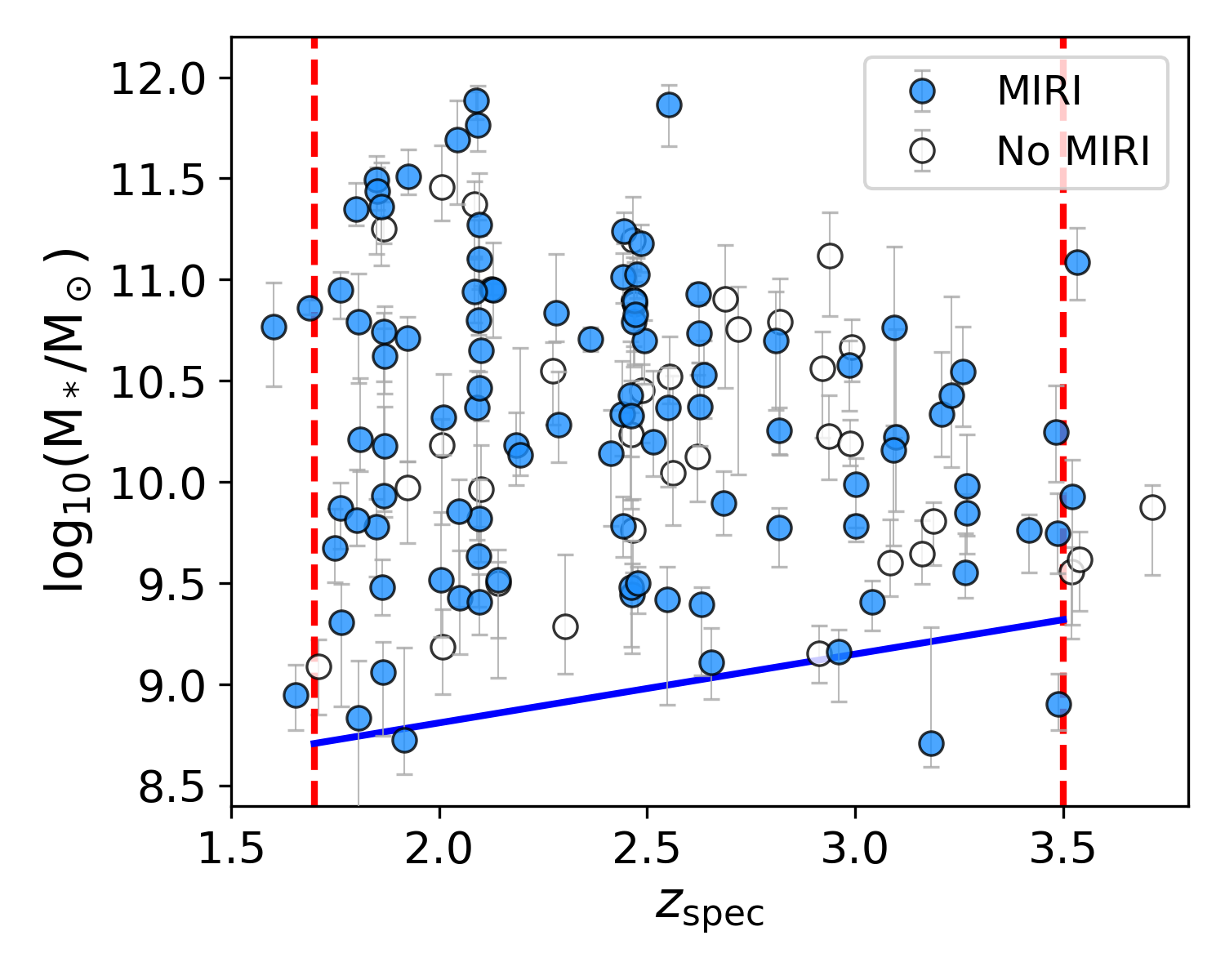}
    \caption{Stellar mass vs. spectroscopic redshift of the Blue Jay sample. Filled circles represent galaxies observed by MIRI. The blue and red lines indicate the limits of the sample selection for stellar mass and redshifts, respectively. Galaxies lying outside of these boundaries satisfied the selection criteria in the 3D-HST photometric catalogue before precise spectroscopic redshifts and updated stellar masses were available.}
    \label{fig:bluejay_sample}
\end{figure}

The Blue Jay targets were carefully selected from the 3D-HST photometric catalogue \citep{brammer2012,skelton2014,momcheva2016} in an effort to achieve a uniform distribution in redshifts (\mbox{$1.7<z<3.5$}) and stellar masses ~\mbox{($\log (M_*/M_{\odot})\gtrsim9$)}. This ensures a representative sample of galaxies at cosmic noon, spanning the full $z$-$M_\ast$ parameter space without being biased in colour, SFR, or morphology. Figure \ref{fig:bluejay_sample} shows the sample on the mass vs. redshift plane, indicating the redshift cutoffs (red dashed lines) and the mass completeness limit (blue solid line). For 10 out of the 153 Blue Jay targets (~7\%) no spectroscopic redshift measurements are available. Four further sources are high-redshift filler galaxies ($5.7 < z < 7$) selected via a dropout technique and are not shown in Fig. \ref{fig:bluejay_sample}. The complete photometric\footnote{\href{https://doi.org/10.5281/zenodo.17333505}{https://doi.org/10.5281/zenodo.17333505}} and spectroscopic\footnote{\href{https://doi.org/10.5281/zenodo.20080314}{https://doi.org/10.5281/zenodo.20080314}} catalogues are publicly available and details about their contents can be found in \citet{belli2026bluejaysurveydeep}.


\subsection{MIRI Imaging}

We extend the Blue Jay photometric catalogue to the mid-infrared by including archival JWST/MIRI observations. All photometric data used in this work are taken from the Mikulski Archive for Space Telescopes (MAST)\footnote{\href{https://mast.stsci.edu/search/ui/\#/jwst}{https://mast.stsci.edu/search/ui/\#/jwst}} as product-level 3 mosaics produced by the standard JWST calibration pipeline. We gathered data from three major JWST imaging programmes in the COSMOS field: PRIMER (GO 1837; \citealp{dunlop2021primer}) in the F770W and F1800W bands, COSMOS-Web (GO 1727; \citealp{cosmos-web}) in the F770W band, and \mbox{COSMOS-3D} (\mbox{GO 5893}; \citealp{cosmos3d}) in the F1000W and F2100W bands. By matching the coordinates of the Blue Jay sources we find that 111 Blue Jay galaxies have MIRI observations, yielding up to four MIRI filters per galaxy. 

We checked for astrometric offsets by computing centroid positions in the MIRI images and comparing them to a reference NIRCam image in the F444W band, which was astrometrically registered using stellar coordinates from Gaia \citep{belli2026bluejaysurveydeep}. Since the COSMOS field is observable twice per year and the orientation of JWST is constrained by the position of the Sun, imaging programmes obtained during the same visibility window typically show similar systematic offsets. We therefore split each data set into two observing periods and determined the astrometric correction separately for each observation. When available, we derived the offsets using MIRI/F770W relative to NIRCam/F444W; otherwise we used MIRI/F1000W, as these filters lie closest in wavelength. The corrections are computed as the median offsets in RA and Dec between the MIRI and NIRCam centroids, with the median absolute deviation (MAD) adopted as a robust measure of the scatter. We then applied the astrometric shift to all the mosaics of the same observing window, including the longer wavelength bands F1800W and F2100W. The typical systematic shift per observation is $|\Delta|_{med}=0.10-0.25$", with a scatter of $\mathrm{MAD}\leq0.04$" in both coordinates. We verify the robustness of these values by repeating the analysis with the longer wavelength filters (F1800W and F2100W), finding consistent offsets within the MAD. The applied corrections are listed in Table \ref{tab:offsets} in Appendix \ref{app:astrometry}.

\section{MIRI Photometry}    \label{sc:photometry}
Photometry was performed on the MIRI observations using a custom-built photometric pipeline, which is published as the stand-alone Python package \texttt{miriutils}\footnote{\href{https://doi.org/10.5281/zenodo.24204809}{https://doi.org/10.5281/zenodo.24204809}}. This library makes extensive use of the \texttt{photutils} package \citep{bradley2022photutils} and provides a streamlined framework for the production of galaxy cutouts from large mosaics, astrometric corrections, local background modelling, and automated aperture photometry.

\subsection{Definition of Apertures}
\label{ssc:aperture_definition}
To ensure consistency with the Blue Jay photometric catalogue, we adopt the elliptical apertures based on the NIRCam/F444W morphology described in \citet{belli2026bluejaysurveydeep}. We rotated each aperture to match the orientation of the MIRI image for each observing run, and then rescaled the aperture by a factor of two to account for the significantly larger point spread functions (PSFs) observed in MIRI images compared to NIRCam. For galaxies with nearby companions where this would result in significant blending, we adopted a custom scaling factor between 1.0 and 2.0. 

Aperture corrections were applied to account for flux falling outside the elliptical aperture due to the MIRI PSF, assuming that the source is point-like. The corrections were derived separately for each filter using PSF models generated with the \texttt{STPSF} package (formerly \texttt{WebbPSF}; \citet{perrin2012stpsf}). We used detector-sampled PSF models that include instrumental effects such as geometric distortion and inter-pixel capacitance. For each band and aperture size, the correction factor was computed as the ratio between the total normalised PSF flux and the flux enclosed within the aperture,
\begin{equation}
    \alpha_{\mathrm{corr}} = \frac{f_{\mathrm{total,PSF}}}{f_{\mathrm{aperture,PSF}}}.
\end{equation}

To validate the rescaling, we compared the aperture-corrected fluxes obtained within the original apertures (derived from NIRCam images) and within the rescaled apertures. The rescaled apertures consistently yield higher fluxes across all bands, with median ratios ($S_{large}/S_{small}$) increasing from 1.39 in F770W to 1.92 in F2100W. We also performed curve-of-growth analysis and found that the measured flux only reaches a stable plateau at radii significantly larger than those defined by the original NIRCam apertures. This behaviour indicates that the galaxies are spatially resolved relative to the MIRI PSF. Consequently, point-source aperture corrections applied to the smaller apertures underestimate the total flux, justifying the use of larger apertures to better recover the integrated emission.

\subsection{Background Modelling and Aperture Photometry}
\label{ssc:background_and_aperture_phot}

The sensitivity of the MIRI mosaics is largely limited by the spatially varying background emission, particularly at wavelengths beyond 10 $\upmu$m where thermal emission from the observatory contributes significantly. Accurate background estimation is therefore critical. Several studies have used custom adaptations of the JWST calibration pipeline to model the background of MIRI images \citep[e.g.][]{harish2025reduction,alberts2024smiles}, but in this work we follow a simpler cutout-based approach applied directly to fully calibrated level-3 mosaics.

\begin{figure*}[t]
    \centering
    \includegraphics[width=0.9\linewidth]{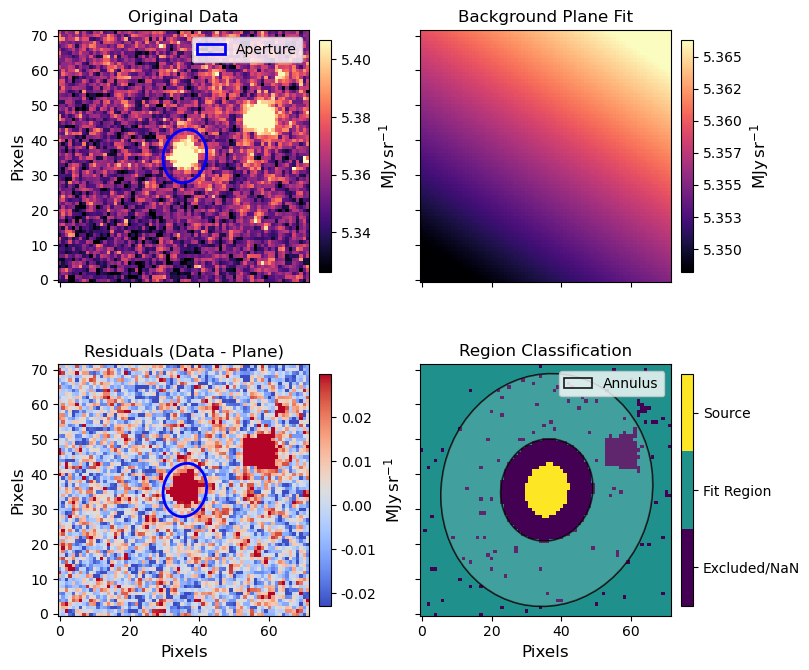}
    \caption{Overview of the regions used for background modelling and aperture photometry. Top row: original image (left) and the two-dimensional background plane fit (right). Bottom row: background-subtracted image (left) and region map of the different regions used (right). The yellow region marks the source aperture and background statistics are taken from the elliptical annulus. Excluded pixels are shown in dark blue.}
    \label{fig:phot_regions}
\end{figure*}

For each galaxy we extract a cutout centred on the source. The procedure follows an iterative approach to detect and measure the background emission, which is illustrated in Fig.~\ref{fig:phot_regions} and includes the following steps:
\begin{enumerate}
    \item Source and Neighbour Masking: We first mask the target galaxy using the elliptical aperture enlarged by a buffer region to prevent the wings of the source profile from biasing the background estimate. We perform initial sigma-clipping at 2.5 standard deviations on the remaining data to remove bright pixels. Sources are then identified with the \texttt{detect\_sources} function, using a detection threshold defined as the median background level plus two times the background standard deviation. This procedure allows us to identify nearby companions and faint extended emission by large sources, which are added to the final masking map.
    
    \item Plane Fitting: To account for large-scale gradients within the background flux, we fit a 2D plane to the unmasked pixels. For the short-wavelength bands ($\lambda \leq 14\upmu$m), we use a standard least-squares fit. However, large-scale thermal gradients and detector artefacts present at longer wavelengths introduce spatial structure that is poorly described by a uniform background model. We therefore implement a Weighted Least Squares (WLS) approach for the F1800W and F2100W bands. We apply a Gaussian weighting scheme centred on the target with a scale length of 2.5 times the aperture semi-major axis. This prioritises the local background near the source over distant pixels and avoids background over-subtractions due to the presence of detector artefacts.
    
    \item Iterative Clipping: The plane is fit over several iterations. In each step, we calculate the residuals and perform sigma-clipping to remove previously undetected outliers. This process is repeated up to three times, which is usually sufficient for the pixel mask to converge.
\end{enumerate}

The final plane is then subtracted from the science data. We compute the background modelling uncertainty $\sigma_{bkg}$ by taking the MAD of the residuals within an elliptical annulus around the source. We performed aperture photometry using \texttt{aperture\_photometry} from the Photutils package with \texttt{method="exact"}, ensuring that aperture pixels are weighted by their fractional overlap. We then define two types of error estimates:

\begin{itemize}
    \item Propagated Errors: This term, $\sigma_{prop}$, accounts for statistical uncertainty, high-frequency detector noise, and uncertainty in the local sky subtraction. It is defined by the quadrature sum over all the pixels within the aperture:
    \begin{equation}
    \label{eq:sigma_prop}
        \sigma_{\text{prop}} = \sqrt{\sum_{i \in \text{ap}} \sigma_{\text{err},\,i}^2 + \left( \sigma_{\text{bkg}} \cdot \sqrt{N_{\text{pix}}} \right)^2}\,,
    \end{equation}
    where $\sigma_{err,i}$ is the per-pixel uncertainty from the instrument's \texttt{ERR} extension, $\sigma_{bkg}$ is the robust background noise, and $N_{\mathrm{pix}}$ is the number of pixels within the aperture.

    \item Empirical RMS: We performed a "blank aperture" test by measuring the flux in 200 apertures placed randomly in empty sky regions on the source-masked science data. The empirical uncertainty $\sigma_{\text{emp}}$ is defined as the MAD of these measurements. This method captures the effects of pixel-to-pixel correlation and large-scale background fluctuations that are not entirely captured by the 2D-background model.
\end{itemize}

The uncertainty reported in the photometric catalogue is chosen as $\sigma_{final}=\mathrm{max}(\sigma_{prop},\sigma_{emp})$. This approach ensures that our errors remain robust even at longer wavelengths, where detector artefacts and thermal gradients are more common. Finally, all fluxes and uncertainties were scaled by the wavelength-dependent aperture correction factor computed with the \texttt{STPSF} library, as described in Section~\ref{ssc:aperture_definition}.

\subsection{Validation of the Photometric Measurements}
\label{ssc:validation_of_photometry}

To validate the robustness of our photometric pipeline, we compare our measurements with the COSMOS2025 photometric catalogue published in the COSMOS-Web Public Data \mbox{Release 1} (DR1; \citealp{cosmos2025}). Since COSMOS-Web only used the F770W band, we limit our comparison to this filter. We find 58 Blue Jay targets in the COSMOS-Web catalogue by matching their celestial coordinates, adopting a maximum angular separation of 0.3 arcseconds (the number drops to 55 if the maximum separation is tightened to 0.2 arcseconds). We used the columns "\texttt{flux\_auto\_f770w}" and "\texttt{flux\_err\_auto\_f770w}" to obtain the aperture-corrected fluxes and corresponding measurement errors from the COSMOS2025 catalogue.

\begin{figure}[t]
    \centering
    \includegraphics[width=\hsize]{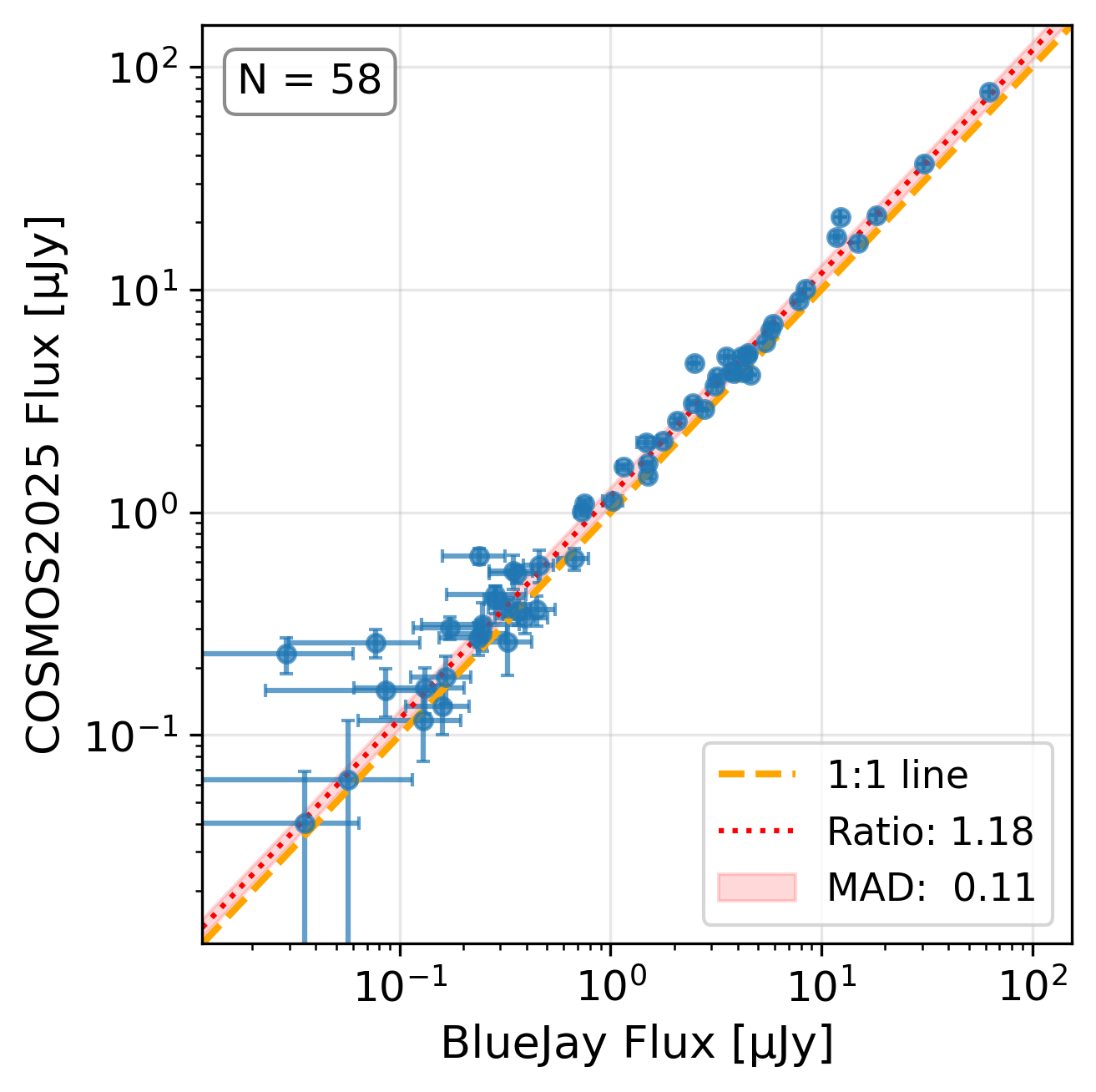}
    \caption{Flux comparison in the F770W band between this work and the COSMOS2025 photometry from \citet{cosmos2025}. Brighter sources cluster closely around the 1:1 ratio (orange line), with a Pearson correlation coefficient of $r=0.996$ for a total of 58 sources. We obtain a median flux ratio of $1.182 \pm 0.110$ between the COSMOS2025 and the Blue Jay sample, which is used as a calibration for our measurements.}
    \label{fig:cosmos2025fluxes}
\end{figure}

The flux comparison is shown in Fig.~\ref{fig:cosmos2025fluxes}. We generally find excellent agreement, particularly for brighter systems. The increased scatter towards the faint end remains largely consistent with the quoted photometric uncertainties and likely reflects differences in the adopted background modelling procedures. However, looking more closely at the comparison, we find a systematic bias between the two catalogues, with COSMOS2025 fluxes lying $\sim18$\% above the Blue Jay measurements, with a modest scatter of $\pm11$\%. This offset is a likely result of the point-source aperture corrections discussed in Section \ref{ssc:aperture_definition}, which do not properly account for the extended light profiles of galaxies but instead assume a point source profile. In contrast, COSMOS2025 employs a combination of \texttt{PSFEx}\footnote{\href{https://github.com/astromatic/psfex}{https://github.com/astromatic/psfex}} \citep{bertin2011psfex} and \texttt{SourceXtractor++} \citep{kummel2022sourcextractor} to obtain highly accurate PSF models.

Independently, we fit the HST, NIRCam, and MIRI photometry with the Bayesian inference code \textsc{Prospector}, using the setup described in \ref{ssc:sed_model_setup}. We find that even when the MIRI photometry is included in the fit, the best-fit model spectra overpredict the flux at 7.7$\upmu$m by 28\% with a MAD of ~25\%. We conclude that our measurements are indeed biased low in such a way that the code is unable to reproduce the low fluxes observed. We thus calibrate all reported fluxes using the COSMOS2025 photometric catalogue by applying a scaling factor of 1.182 to all measurements. We also add an 11\% relative uncertainty in quadrature when computing the propagated errors in Eq.~\eqref{eq:sigma_prop}, corresponding to our uncertainty on the flux ratios in Fig.~\ref{fig:cosmos2025fluxes}.

\subsection{The Blue Jay MIRI Photometric Catalogue}
\label{ssc:catalogue}

We make the Blue Jay MIRI Photometric Catalogue publicly available\footnote{\href{https://doi.org/10.5281/zenodo.24204865}{https://doi.org/10.5281/zenodo.24204865}}. The catalogue contains aperture-corrected fluxes, associated uncertainties, background measurements, and quality-control flags for all available MIRI filters.
A full description of the catalogue columns and the quality flagging routine is provided in Appendix~\ref{app:phot_table_description}.

\begin{figure}[t]
    \centering
    \includegraphics[width=\hsize]{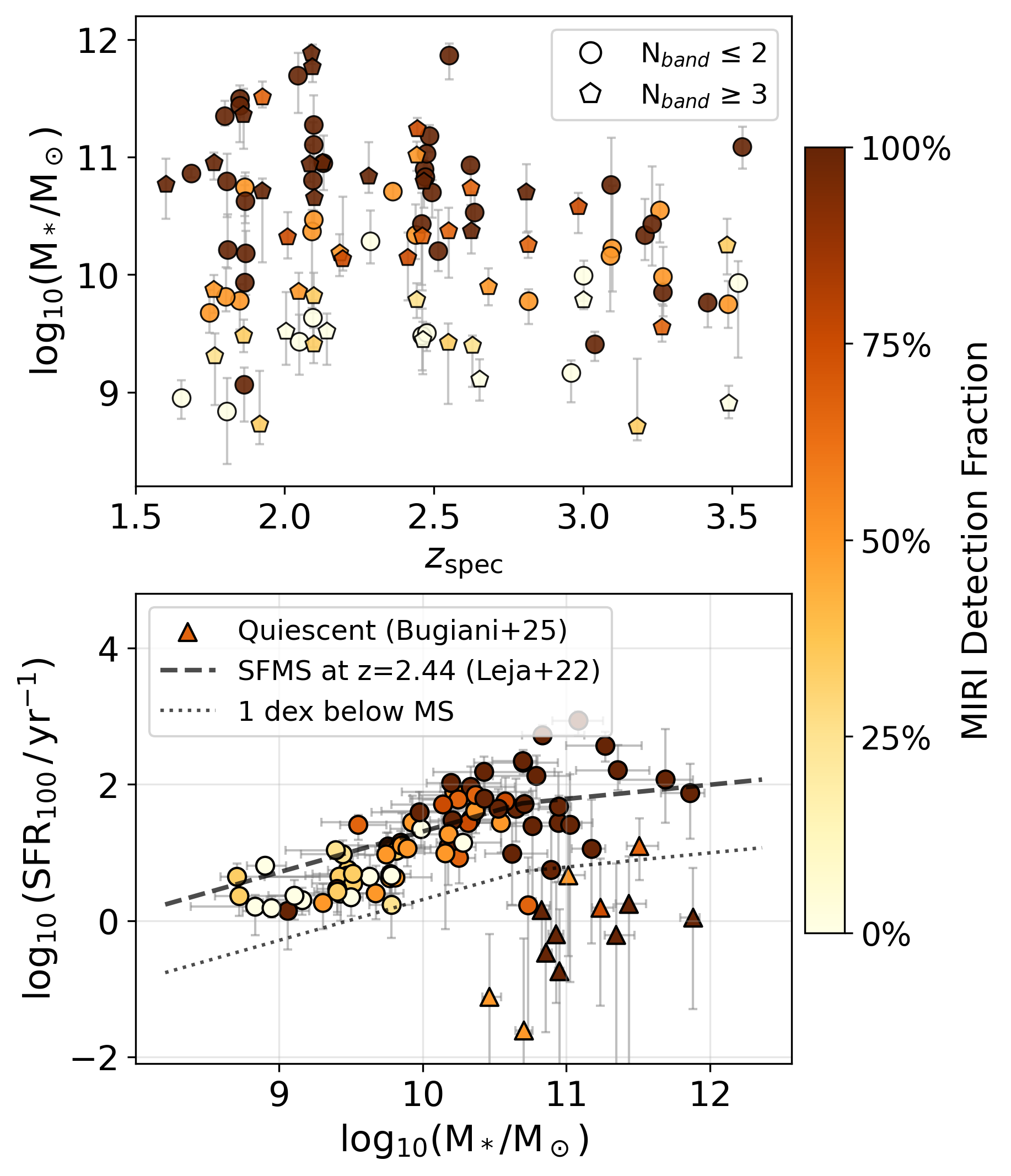}
    \caption{Top panel: Stellar mass–redshift distribution of the Blue Jay MIRI sample. Circles denote galaxies with fewer than three available MIRI bands, while pentagons represent galaxies with three or more available bands. Bottom panel: The MIRI sample with respect to the star-forming main sequence (SFMS) of \cite{leja2022sfms} at the median sample redshift ($z=2.44$). Triangles denote the quiescent galaxies confirmed in \citet{bugiani2025feedback}. The colour scale indicates the fraction of available MIRI bands in which each galaxy is detected. We find that stellar mass is a stronger predictor for MIRI detections than the SFR within the last 100 Myr.}
    \label{fig:sample_detections}
\end{figure}

We find that 111 out of 153 Blue Jay galaxies are observed in at least one MIRI band, hereafter referred to as the MIRI sample. For each available MIRI image, we compute the signal-to-noise ratio (SNR) and mark galaxies as detected if they exceed a threshold of $\mathrm{SNR} > 3$. For each galaxy, we calculate the detection fraction, which is defined as the number of MIRI images in which the source is detected divided by the total number of bands available. In the top panel of Fig.~\ref{fig:sample_detections} we show the $M_\ast$--$z$ plane for the MIRI sample in analogy to Fig.~\ref{fig:bluejay_sample}. We find that the detection success rate is strongly correlated with stellar mass: the most massive galaxies ($\mathrm{log(M_\ast/M_{\odot})}\gtrsim10$) are almost universally detected in all available bands (detection fraction $\approx100\%$), showing only a slight dependence on redshift. Crucially, many of the galaxies observed in three or more MIRI bands (pentagonal markers) are reliably detected, providing valuable data to investigate dust energy balance. In the bottom panel of Fig.~\ref{fig:sample_detections} we show the relative position of the MIRI sample with respect to the star-forming main sequence (SFMS) by \citet{leja2022sfms} at the median sample redshift ($z=2.44$), using the SFR averaged over the last 100 Myr ($\mathrm{SFR_{100}}$). At these redshifts the bluest MIRI bands (F770W and F1000W) still probe the rest-frame near-infrared stellar continuum, enabling detections of massive galaxies even in systems with low dust content or suppressed star formation activity. A comprehensive overview of each galaxy's detection status is provided in Fig.~\ref{fig:miri_stats} in Appendix~\ref{app:detection_stats}.


\section{Investigating Dust Energy Balance}
\label{sc:investigating_dust_energy_balance}

The main goal of this study is to test the dust energy balance assumption by comparing the predicted and measured MIRI photometry. The predicted MIRI emission is derived from SED fits to the HST and JWST/NIRCam photometry, and is thus based solely on the inferred stellar populations and dust attenuation at shorter wavelengths, under the assumption of dust energy balance. By investigating how accurately the models reproduce MIRI observations and by examining how the residuals depend on galaxy properties, we assess the general reliability of the dust energy balance framework as a simplified description of dust reprocessing, implemented in many modern SED fitting codes.

Of the 111 Blue Jay galaxies observed with MIRI at cosmic noon, we exclude 8 objects due to the lack of a spectroscopic redshift measurement. We also exclude 10 galaxies for which the aperture could not be sufficiently rescaled due to the presence of detector artefacts or companions in the cutouts. Lastly, we remove two high-redshift filler targets from the sample, as well as two galaxies hosting active galactic nuclei (AGNs) with confirmed broad lines, since our SED model does not include an AGN component. The final sample thus consists of 89 Blue Jay galaxies at Cosmic Noon.

\subsection{SED Fitting}
\label{ssc:sed_model_setup}

The spectral energy distributions were modelled with \textsc{Prospector} \citep{johnson2021prospector}, a Bayesian stellar population inference code based on the \textsc{Fsps} library \citep{conroy2009fsps}. We follow the modelling approach of \citet{park2024widespread} and \citet{bugiani2025feedback}, adopting a \citet{chabrier2003imf} initial mass function (IMF) and allowing for variable stellar metallicity with a prior range of $-2 < \log(Z_\odot) < 0.5$. We use non-parametric star formation histories (SFHs) by defining 14 age bins that are fixed to $0$, $5$ Myr and 10 Myr, followed by an even logarithmic spacing to the maximum lookback time. The SFR ratios between adjacent bins follow a Student's $t$ continuity prior with a scale width of 0.3 dex and two degrees of freedom ($\nu=2$), which mildly penalises sharp transitions in the star formation rates of adjacent bins \citep{leja2019nonparametric}.

We model dust attenuation assuming a two-component \citet{charlot2000dust} model, consisting of a diffuse dust component affecting all stars and an additional birth-cloud component surrounding young stellar populations. The model is characterised by three free parameters: The diffuse attenuation follows a modified \citet{calzetti_2000_dust} law parametrised by the optical depth at $5500\,\text{\AA}$ ($\tau_{V,2}$), which is modulated by a power-law index $n$ following \citet{kriek2013dust}. Stars younger than $10\,\text{Myr}$ are attenuated by an additional optical depth expressed as a fraction of the diffuse dust ($\eta = \tau_{V,1} / \tau_{V,2}$). A thorough description of the individual parameters is given in \citet{maheson2025bigdusty}. 

Dust emission is included under the assumption of energy balance, where the total infrared luminosity matches the energy absorbed by dust ($L_{\mathrm{IR}} = L_{\mathrm{abs}}$). We make use of the \citet{draine2007dust} templates with three free parameters controlling the mass fraction of polycyclic aromatic hydrocarbons ($q_{\mathrm{PAH}}$), the lower threshold of the starlight intensity distribution $U_{min}$, and the fraction $\gamma$ of dust exposed to elevated radiation fields ($U>U_{\mathrm{min}}$). We adopt uniform top-hat priors for all parameters: $q_{\mathrm{PAH}} \in [0.5,\, 10.0]\,\%$, $U_{\mathrm{min}} \in [0.1,\, 25.0]$, and $\gamma \in [0.0,\, 1.0]$. We note that empirical constraints from resolved nearby galaxies favour narrower prior distributions \citep{leja2017deriving, leja2019older_more_quiescent}. However, these calibrations are subject to sample selection effects and may not accurately represent the ISM conditions of galaxies at higher redshifts. Adopting wide uniform ranges thus allows us to test the extrapolation of the rest-frame UV-optical emission using energy balance against MIRI photometry without artificially restricting the mid-infrared parameter space. Lastly, we do not include an AGN component in the SED models since we removed confirmed AGNs from the final sample.

Nebular line and continuum emission are modelled self-consistently with the ionising photon budget of the stellar populations using \textsc{Cloudy} \citep{ferland1998cloudy, ferland2013cloudy}. For each galaxy, the redshift is assigned a narrow Gaussian prior centred on the value of the spectroscopic redshift with $\sigma=0.005$. At this stage, we only include photometry from HST and NIRCam in the \textsc{Prospector} fits, but we use the NIRSpec data to anchor the model to the spectroscopic redshift. We will discuss the effect of including the spectroscopy in the fitting process in Section~\ref{ssc:adding_nirspec}.

\begin{figure}[t]
    \centering
    \includegraphics[width=\hsize]{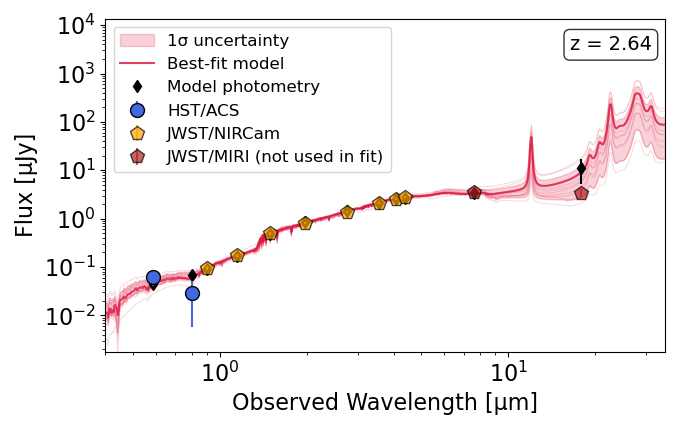}
    \caption{Example of a best-fit model spectrum and ten probability-weighted random posterior samples (red lines). The red-shaded region demarcates the 1$\sigma$ uncertainty envelope. Photometry from HST (blue dots), NIRCam (orange pentagons) and MIRI (red pentagons; overlaid) are shown together with the predicted model photometry (black diamonds). In the F770W filter the model agrees within $\sim5$\% with the observed flux, while the photometry in the F1800W band is overpredicted by a factor of 3.24.}
    \label{fig:sed_example}
\end{figure}

We fit the model to eight JWST/NIRCam bands and two HST/ACS bands. We do not include the HST/WFC3 bands because they fully overlap in wavelength with the NIRCam filters and have a much lower signal-to-noise ratio. An example of a best-fit SED is shown in Fig.~\ref{fig:sed_example}. The MIRI photometry, shown in the figure, is not included in the SED fit. We therefore manually computed the predicted MIRI fluxes by convolving posterior model spectra with the corresponding filter transmission curves. For each band, uncertainties were estimated from the 16th and 84th percentiles of 200 posterior samples, so that
\begin{equation}
    \sigma_{model} = \frac{f_{model}^{84th} - f_{model}^{16th}}{2}\,.
\end{equation}
In the example shown in Fig.~\ref{fig:sed_example}, the F770W flux agrees within $\sim5\,\%$ of the model photometry, whereas the F1800W flux is overpredicted by a factor of 3.24,

\subsection{Comparing predicted and observed photometry}
\label{ssc:systematic_trends_with_wavelength}

We first quantify the agreement between the model and the MIRI observations by computing the logarithmic flux ratio
\begin{equation}    
    R_i = \log_{10}\left(\frac{f_{model,\,i}}{f_{obs,\,i}}\right)\,,
    \label{eq:log_ratio}
\end{equation}
where $f_{obs,i}$ and $f_{model,i}$ refer to the observed and model flux in the \textit{i}-th band, respectively. Since this ratio compares the fluxes irrespective of the underlying errors, we limit our analysis to detected sources ($\mathrm{SNR}>3$). 

We find that the median logarithmic offset is positive, implying that the model overpredicts the observations, particularly at longer wavelengths, where the discrepancy increases from 0.04~dex in F770W to 0.47~dex in F2100W. The close agreement within $\sim$10\% in the F770W band is expected, as it lies close to the wavelength range directly constrained by NIRCam and HST data, where the stellar continuum still dominates the emitted flux. In contrast, the F2100W flux depends almost entirely on the amount of absorbed starlight that is translated into dust emission via dust energy balance, resulting in a median model overprediction of a factor of $\sim3$.

\begin{figure*}[t]
    \centering
    \includegraphics[width=\linewidth]{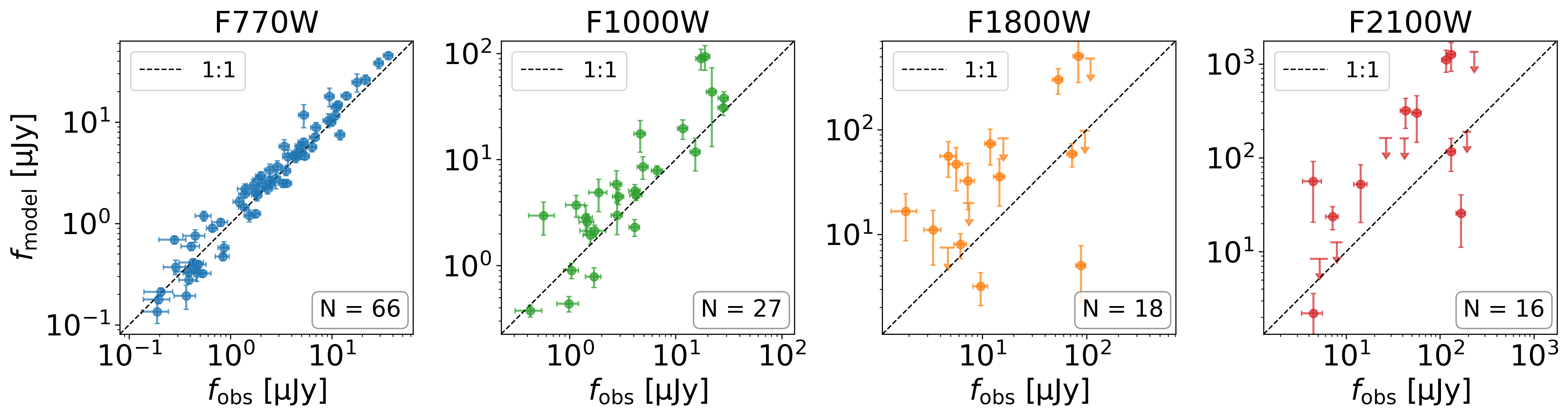}
    \caption{Comparison of observed ($f_{\mathrm{obs}}$) and predicted MIRI fluxes ($f_{\mathrm{model}}$) in each filter. The number of galaxies in each band is shown in the bottom right corner and the grey dashed line indicates perfect agreement. Upper limits are displayed in cases where the model errors are larger than the flux. The model consistently overpredicts the observed fluxes in all MIRI filters, with median discrepancies increasing from around 10\% in F770W to a factor of $\sim3$ in F2100W.}
    \label{fig:obs_vs_model_log}
\end{figure*}

\begin{figure*}[!h]
    \centering
    \includegraphics[width=0.9\linewidth]{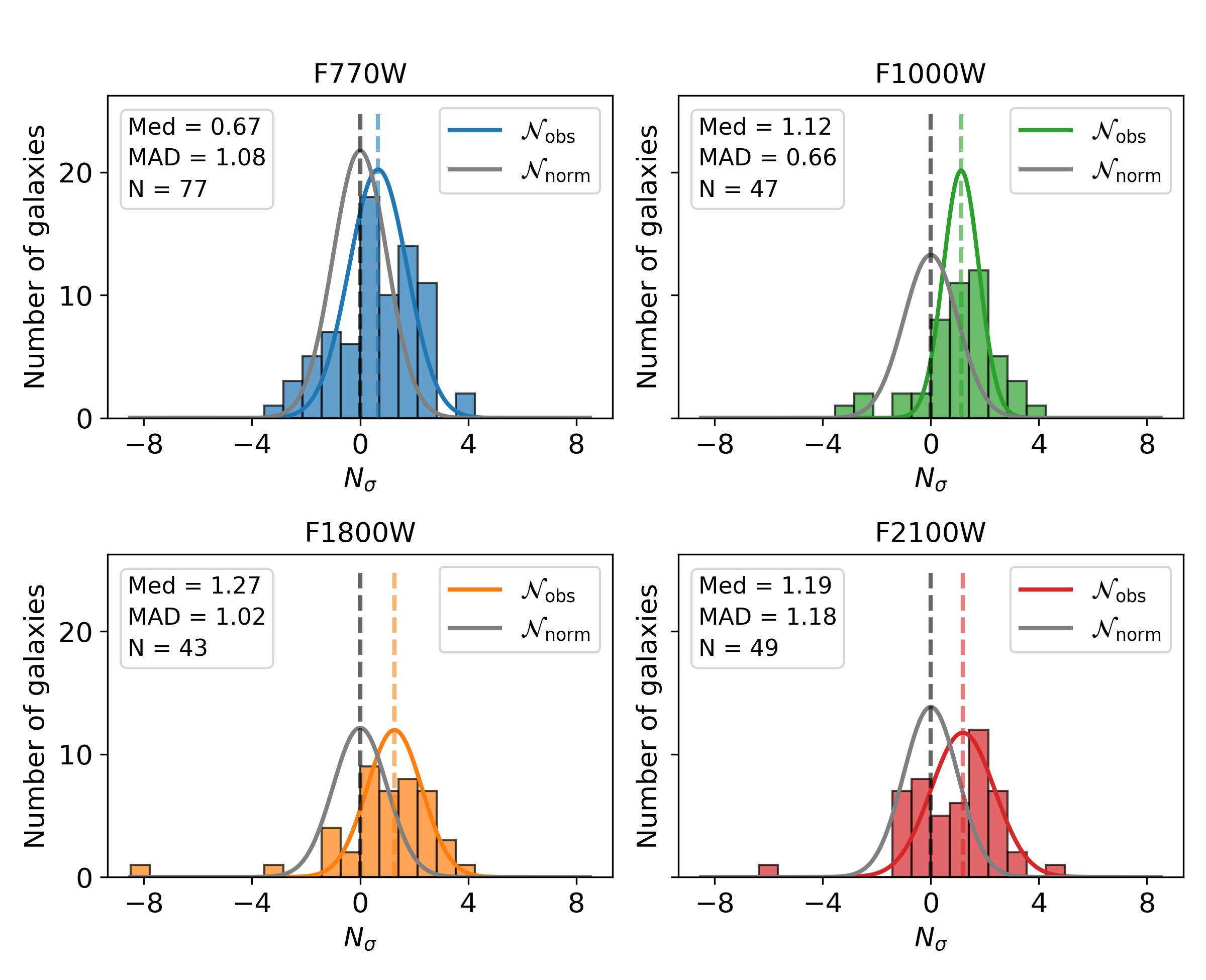}
    \caption{Distribution of normalised residuals ($N_\sigma$) for each of the four MIRI bands, as defined in Eq.~\ref{eq:nsigma}. The median, MAD, and number of galaxies are shown in the top left corner of each panel. A normalised reference Gaussian distribution $\mathcal{N}_{\mathrm{norm}}$ is shown as a grey line and empirical Gaussian fits to the distributions are displayed in the corresponding colour. Vertical dashed lines indicate zero and the median offset of the distribution. All distributions are positively offset from the reference Gaussian, indicating an overprediction of mid-infrared fluxes.}
    \label{fig:nsigmas}
\end{figure*}

The wavelength-dependent nature of the overpredictions is illustrated in Fig.~\ref{fig:obs_vs_model_log}, which shows a comparison between observed and predicted fluxes in the four MIRI bands. While the F770W filter (left-most panel) shows a relatively tight correlation around the 1:1 line, a significant upward shift becomes apparent at longer wavelengths. This trend is accompanied by an increased scatter, measured via the MAD, from 0.09~dex (F770W) to 0.51~dex (F2100W). In all panels, we plot arrows where the error bar is larger than the model flux. We find that the model uncertainties in the longer wavelength bands (F1800W and F2100W) are substantially larger than the photometric errors, indicating that the posterior distributions of the dust emission parameters are broad and largely resemble the adopted priors discussed in Section \ref{ssc:sed_model_setup}.

An alternative way to quantify this offset is to account for the total uncertainty by calculating the normalised residuals:
\begin{equation}
    N_{\sigma,i} = \frac{f_{model,\,i}-f_{obs,\,i}}{\sqrt{\sigma_{model,\,i}^2 + \sigma_{obs,\,i}^2}}\,,
    \label{eq:nsigma}
\end{equation}
where $\sigma_{obs,i}$ and $\sigma_{model,i}$ represent the uncertainties of the observation and model in the $i$-th band, respectively. This definition allows us to compute $N_\sigma$ for both detected and undetected systems. The resulting distributions in each filter are shown in Fig. \ref{fig:nsigmas}. If the code predicted the observed fluxes perfectly, the $N_\sigma$ distribution should follow a normalised Gaussian $\mathcal{N}_{\mathrm{norm}}$. Instead, the curves are systematically shifted towards positive values of $N_\sigma$ in all MIRI bands, again indicating an overprediction of mid-infrared fluxes. The median value of $N_\sigma$ gradually shifts from 0.67 in F770W to 1.19 in F2100W despite the significantly larger model uncertainties at longer wavelengths, which inflate the denominator in Eq.~\ref{eq:nsigma} and keep the distributions close to zero. 

\subsection{Trends with physical properties}
\label{ssc:drivers_of_model_inaccuracy}

We investigate the dependence of the model's predictive power on two important galaxy properties derived from the \textsc{Prospector} fits: the specific star formation rate averaged over the past 100~Myr ($\mathrm{sSFR_{100}}=\mathrm{SFR_{100}}/M_\ast$) and the diffuse dust attenuation ($A_V = 1.086 \cdot \tau_{\nu,2}$). In both cases, we draw 100 posterior-weighted samples from the parameter's MCMC chains and compute the median, 16th and 84th percentiles. We limit our analysis to star-forming galaxies, excluding 12 quiescent galaxies selected in \cite{bugiani2025feedback} that were observed by MIRI, as well as two galaxies that lie more than $\sim$1 dex below the star-forming main sequence shown in the bottom panel of Fig.~\ref{fig:sample_detections}.

\begin{figure}[t]
    \centering
    \includegraphics[width=\hsize]{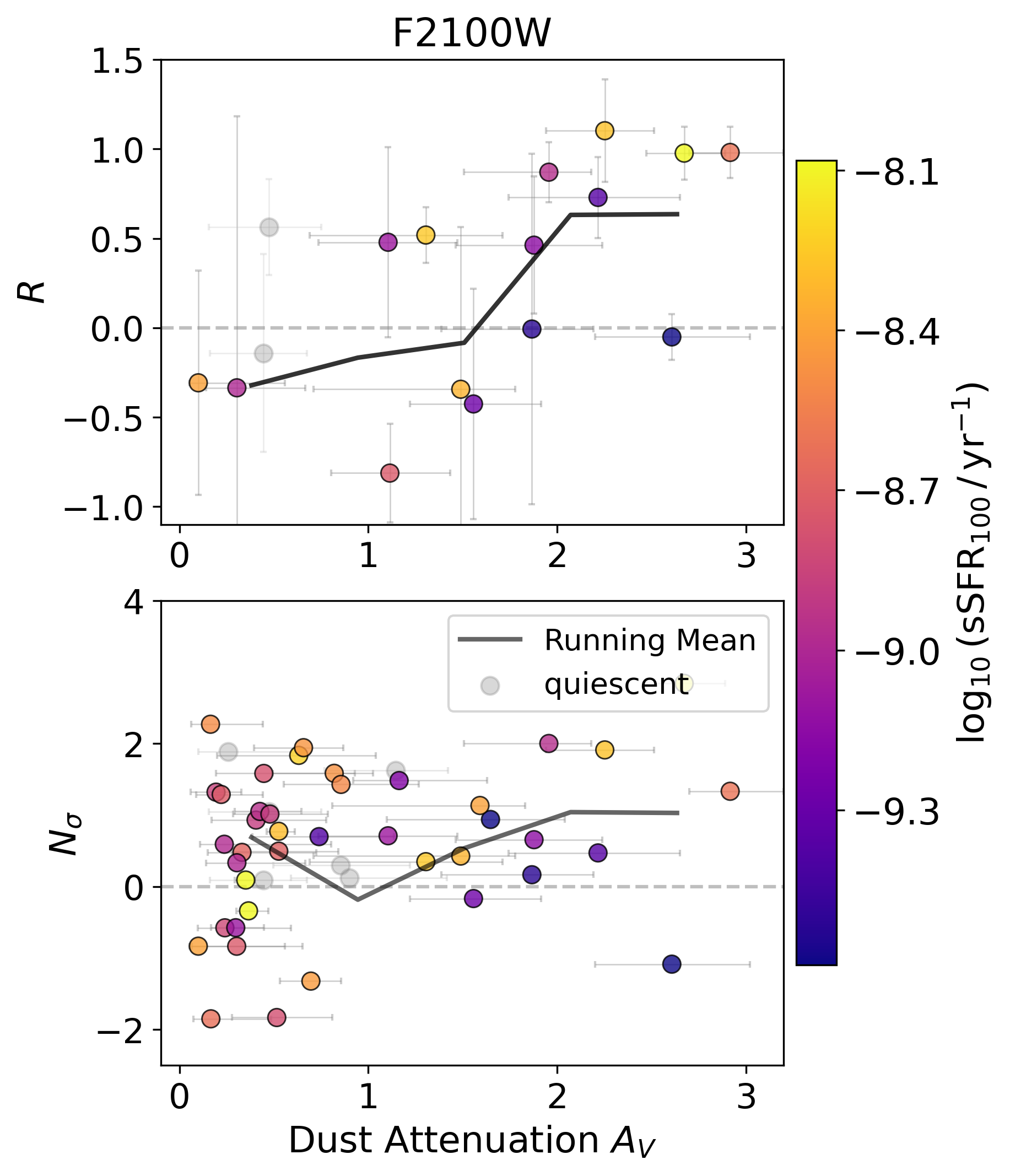}
    \caption{Model prediction accuracy in the F2100W filter as a function of inferred dust attenuation ($A_V$). Top panel: Logarithmic flux ratios $R$ of detected sources ($\mathrm{SNR} > 3$). Bottom panel: Normalised residuals $N_\sigma$, computed for both detected and non-detected galaxies. Data points are colour-coded by specific star formation rate ($\log_{10}(\mathrm{sSFR}_{100})$). Quiescent galaxies are indicated in grey and are excluded from the analysis. The running mean for the star-forming sample is shown as a solid line. In both panels, a pronounced upward trend emerges at $A_V \gtrsim 1.5$, indicating that mid-infrared emission in heavily obscured systems is overpredicted by up to an order of magnitude.}
    \label{fig:f2100w_two_panels}
\end{figure}

In Fig.~\ref{fig:f2100w_two_panels}, we examine the model residuals in the reddest available MIRI band, F2100W, as a function of the inferred dust attenuation $A_V$ and star formation activity $\log(\mathrm{sSFR_{100}})$, shown as colour scale. Both the logarithmic flux ratios ($R$) and the normalised residuals ($N_\sigma$) show a clear dependence on dust obscuration. In low-attenuation systems ($A_V \lesssim 1.0$), the models show no significant systematic offset, with running mean values clustering near $R \approx -0.3$ dex and $N_\sigma \approx 0.7$. In contrast, for heavily obscured galaxies ($A_V \gtrsim 1.5$) the model significantly overpredicts the observed fluxes, with the running mean of $R$ climbing to $+0.63$ dex (corresponding to an overprediction of a factor of $\approx 4.4$), while the running mean of $N_\sigma$ rises mildly to $1.03$. We find a minor trend in which points with lower $\mathrm{sSFR_{100}}$ lie closer to zero than those with intense star formation.

\begin{figure*}[h]
    \centering
    \includegraphics[width=0.95\linewidth]{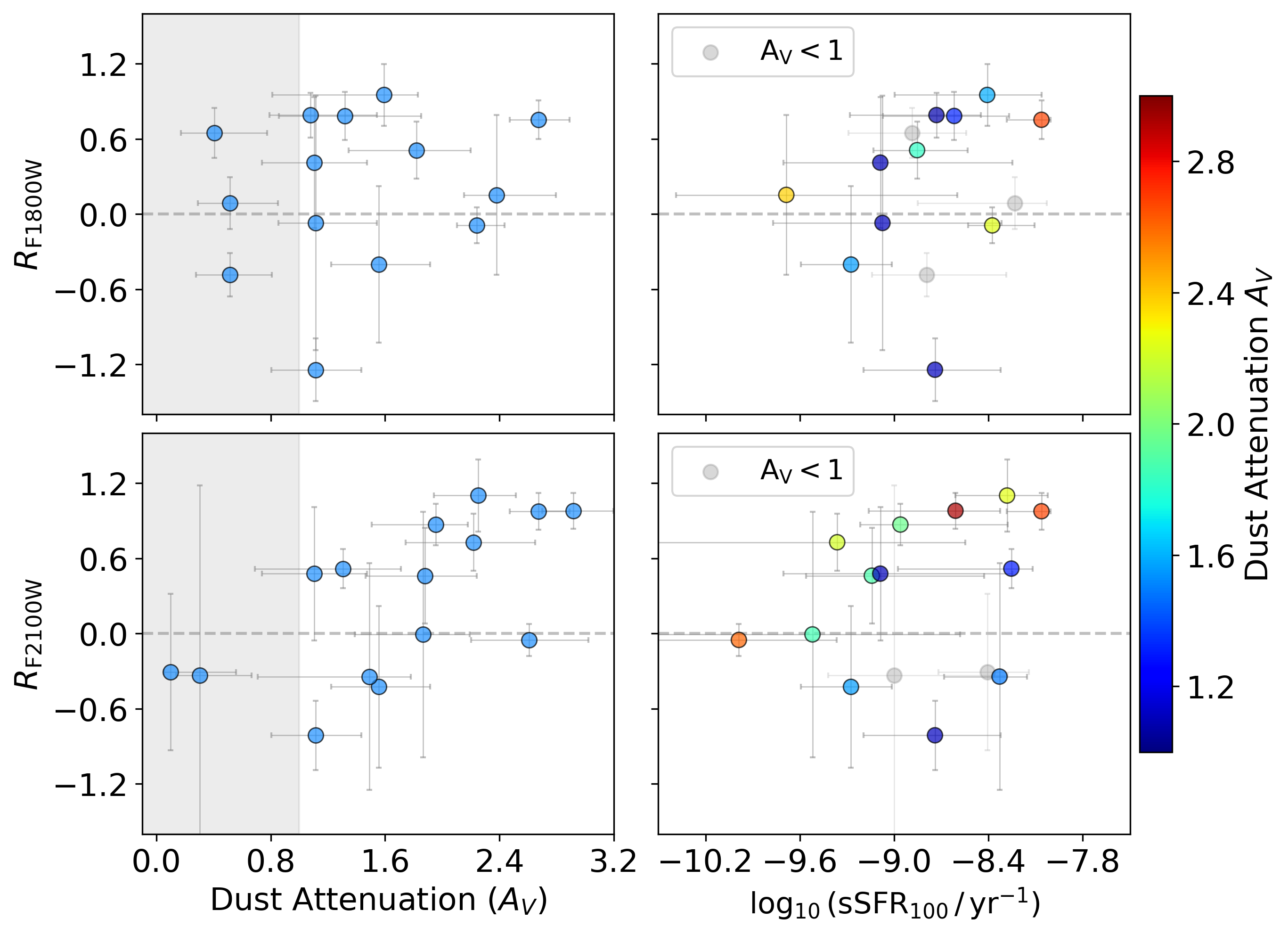}
    \caption{Logarithmic flux ratios between predicted and measured fluxes ($R = \log_{10}({f_{model}/f_{obs}})\,$) in the F1800W (top) and F2100W (bottom) bands. Left panels: $R$ versus dust attenuation $A_V$. The grey-shaded region indicates the low-attenuation regime ($A_V<1$). Right panels: $R$ versus specific star formation rate $\mathrm{sSFR_{100}}$. Points with $A_V>1$ are colour-coded by $A_V$, whilst low-attenuation galaxies are shown in grey. The dashed horizontal lines ($R=0$) indicate perfect agreement.}
    \label{fig:logr_vs_all_params}
\end{figure*}

In Fig.~\ref{fig:logr_vs_all_params}, we examine how the logarithmic flux ratios $R$ scale with dust attenuation $A_V$ (left column) and specific star formation rate $\mathrm{sSFR}_{100}$ (right column) for the two reddest bands, F1800W (top row) and F2100W (bottom row). Mirroring the behaviour seen in Fig.~\ref{fig:f2100w_two_panels}, we find a positive correlation of $R$ with $A_V$ in F1800W, albeit weaker than in F2100W. By contrast, $\mathrm{sSFR}_{100}$ appears to play only a secondary role in determining \textsc{Prospector}'s modelling behaviour: whilst galaxies with higher $\mathrm{sSFR}_{100}$ exhibit slightly larger mid-infrared flux overpredictions, the correlation is considerably weaker and displays larger scatter than the trend with $A_V$.

Together, these findings paint a clear picture in which dust energy balance performs well in relatively unobscured, moderately star-forming galaxies, but systematically overshoots mid-infrared emission in dusty galaxies ($A_V \gtrsim1$). One possible explanation is that the total amount of absorbed starlight ($L_{\mathrm{abs}}$) is overestimated when fitting UV--optical photometry alone. A viable alternative explanation is that the absorbed energy is emitted at longer wavelengths, which cannot be probed with MIRI. This is expected if dusty systems experience significant self-shielding: while higher dust attenuation implies a larger amount of absorbed starlight ($L_{\mathrm{abs}}$), the increased optical depth can act as a shield for the interior dust from intense radiation fields, leading to a lower equilibrium dust temperature throughout the galaxy. As a consequence, most of the absorbed energy would be predominantly re-emitted in the far-infrared dust peak instead of the mid-infrared continuum, which is sensitive to warmer dust components.

\subsection{Rest-frame Optical Constraints from NIRSpec}
\label{ssc:adding_nirspec}

The systematic overprediction of IR fluxes by the best-fit SED model suggests that the dust properties are poorly constrained. We investigate this possibility by fitting the NIRSpec spectroscopy at 1-5$\upmu$m taken by the Blue Jay survey, together with the photometry from HST and NIRCam. This spectro-photometric dataset provides full coverage of the rest-frame optical-UV wavelength range, as well as detailed information on individual absorption lines. The inclusion of medium-resolution spectra requires a more complex setup of the \textsc{Prospector} fit, including proper treatment of outlier pixels and imperfect flux calibration. We refer to \citet{park2024widespread} and Bugiani et al. (in prep.) for details on the spectro-photometric fits to the Blue Jay sample. 

\begin{figure*}[t]
    \centering
    \includegraphics[width=\linewidth]{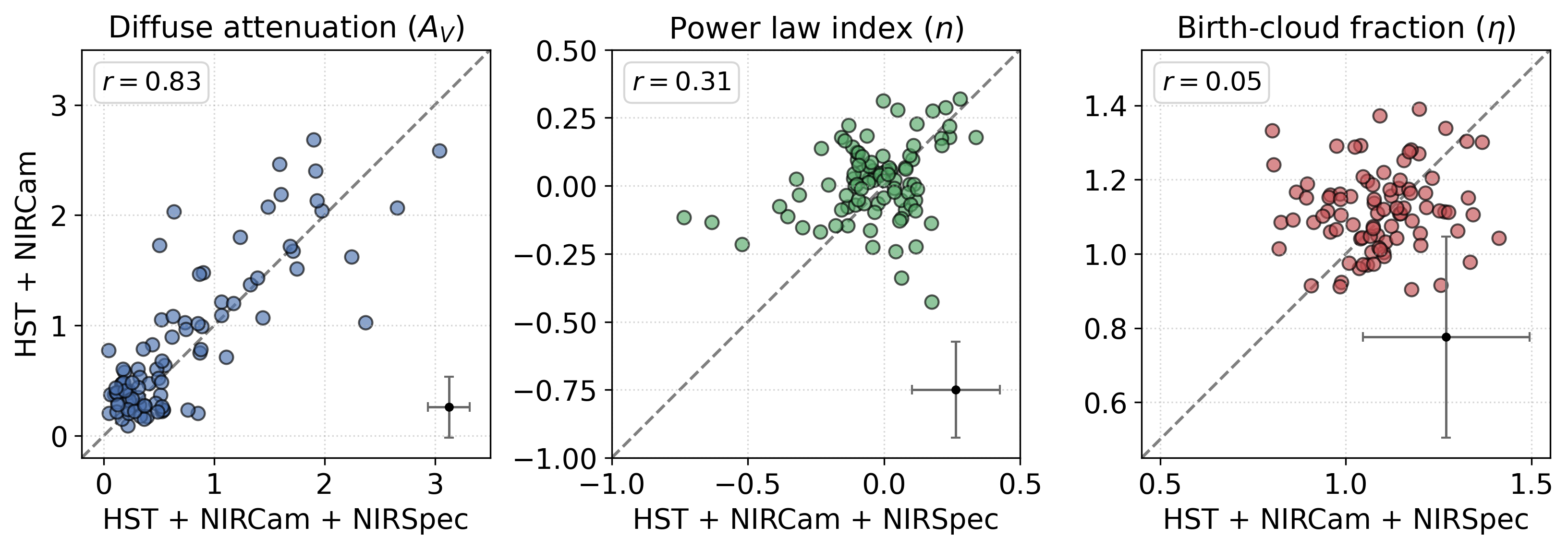}
    \caption{Comparison of the median dust absorption parameters of the \citet{charlot2000dust} dust model when fitting photometry from HST and NIRCam with and without NIRSpec spectroscopy. The Pearson correlation coefficients are shown in the top left corner of each panel, while the median uncertainties are displayed in the bottom right corner.}
    \label{fig:dust_params_nirspec}
\end{figure*}

We compare the inferred dust absorption parameters of SED fits with and without NIRSpec spectroscopy and show the results in Fig. \ref{fig:dust_params_nirspec}. Rather than using a single maximum a posteriori (MAP) value, we report the posterior medians for each parameter, with uncertainties corresponding to the 16th and 84th percentiles derived from the posterior probability distributions. Apart from a moderate scatter, we find good agreement in the diffuse dust attenuation ($A_V$; left panel) when including NIRSpec spectroscopy: dusty galaxies identified with broadband photometry alone remain dusty when spectroscopy is added to the fit. In contrast, the power law index ($n$; middle panel) and the birth-cloud fraction ($\eta \equiv \tau_{V,1}/\tau_{V,2}$; right panel) only show a weak correlation between the two configurations ($r = 0.30$ and $r = 0.05$, respectively), suggesting that NIRSpec allows \textsc{Prospector} to move from average assumptions towards galaxy-specific values.

The physical validity of the updated parameter values is demonstrated by the vast improvement in the predictive power of the resulting SED models, which exhibit significantly lower mid-infrared fluxes when also fitting spectroscopy. Even though no MIRI photometry was used in the SED fit, the median logarithmic residuals ($R$) decrease substantially across all MIRI bands, with the biggest change seen in F2100W. Here, the median overprediction drops from a factor of $\sim 3$ ($+0.39$ dex) to $\sim 1.5$ ($+0.18$ dex), while the scatter (MAD) decreases from $0.56$ dex to $0.36$ dex. These results indicate that NIRSpec successfully breaks parameter degeneracies between dust, age and metallicity inherent to photometry-only fits, thus providing more accurate dust absorption parameters for star-forming galaxies at cosmic noon.



\subsection{Direct IR and Dust Population Constraints from MIRI}
\label{ssc:adding_miri}

Lastly, we fit the measured MIRI photometry together with the available HST and NIRCam data in the UV--optical, without the spectroscopy from NIRSpec. While combining deep rest-frame optical spectroscopy with mid-infrared photometry can in principle break important degeneracies on dust, age and metallicities, spatially matching narrow NIRSpec micro-shutter slits to aperture photometry from MIRI remains observationally challenging due to the substantially broader MIRI point-spread function (PSF) and resulting aperture mismatches.

In contrast to the NIRSpec spectroscopy, which provides detailed information on the absorption of starlight along the line of sight, mid-infrared observations allow us to place direct constraints on dust emission. In particular, many of the PAH features are redshifted beyond $\sim7$ microns at $z\sim2$ so that they fall into the MIRI broadband filters used in this work. The strengths of these features can be used as tracers for star formation intensity and metallicity, while also providing insights into the structure and properties of the PAH molecules residing in the host galaxy \citep[e.g.][]{peeters2004pahstarformation, hunt2010pahmetallicity, rigopoulou2024pah, shivaei2024pah}. 

We find that SED fits including the MIRI photometry completely remove the systematic offsets present in the optical/NIR-only models. The best-fit fluxes across all four MIRI bands match the observations within $\sim 10\%$, exhibiting minimal residual scatter across the star-forming sample. This improved accuracy is further demonstrated by the distribution of the normalised residuals, $N_\sigma$, which cluster tightly around zero with a MAD of less than $1\,\sigma$ in all MIRI filters. Furthermore, for each SED fit we compute the reduced chi-squared statistic for the available MIRI bands and find a median value of $\chi^2_\nu=1.31$, confirming that the model self-consistently reproduces the observed mid-infrared emission.

\begin{figure}[t]
    \centering
    \includegraphics[width=\linewidth]{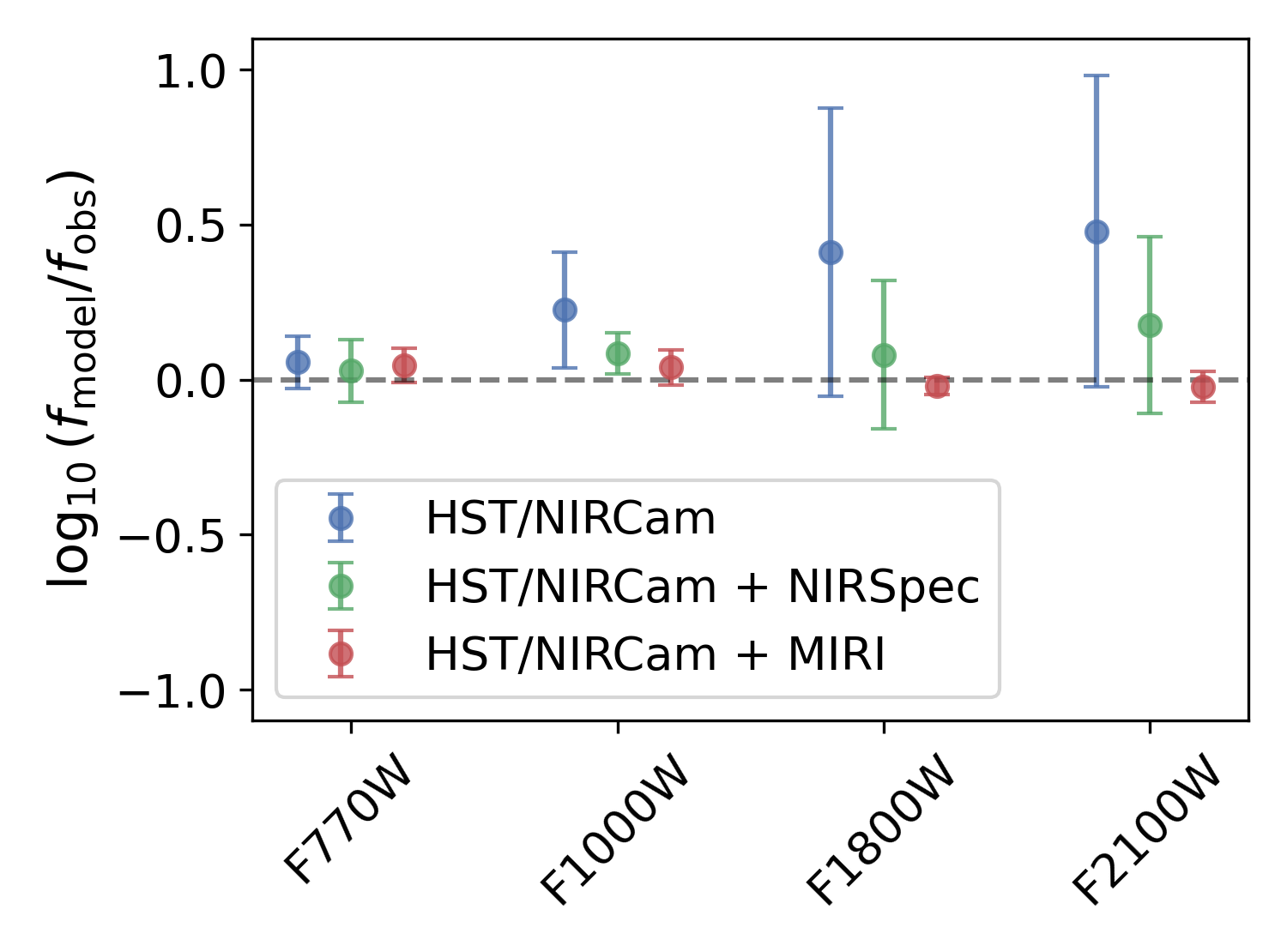}
    \caption{Median and MAD of the logarithmic ratios of predicted and observed fluxes in each MIRI band. We show three fitting configurations: HST + NIRCam photometry (blue), HST + NIRCam photometry + NIRSpec spectroscopy (green), HST + NIRCam + MIRI photometry (Red). The fit quality improves significantly when new data are included in the SED fit, providing either detailed information on dust absorption in the case of NIRSpec spectroscopy, or direct constraints on the IR emission through MIRI photometry.}
    \label{fig:fit_improvement}
\end{figure}

In Fig. \ref{fig:fit_improvement} we compare the logarithmic flux ratios of the best-fit model spectra obtained from three distinct configurations: a fiducial run using only HST + NIRCam photometry (blue), HST + NIRCam + NIRSpec spectroscopy (green), and HST + NIRCam + MIRI photometry (red). For each band, we show the median logarithmic flux ratio and the MAD of the sample. With the exception of the F770W filter, which largely still probes the stellar continuum at $z\sim2$, the model predictions of the MIRI fluxes improve strongly as more data are included in the fit. When fitting only HST and NIRCam photometry, the model is forced to extrapolate the shape of the dust emission based on relatively poor constraints on the total amount of absorbed starlight. This results in significant overpredictions of mid-infrared fluxes together with large uncertainties, particularly at wavelengths redward of 15 microns. Including NIRSpec spectroscopy in the rest-frame UV–optical regime breaks fundamental degeneracies between dust, age, and metallicity, pinning down the true optical depth of absorbed starlight. Crucially, even though NIRSpec is blind to dust emission, this improved characterisation of the absorbed starlight intensity results in substantially improved predictions of the mid-infrared continuum based on dust energy balance, reducing the residuals from a factor of $\sim3$ to about $\sim1.5$.



\begin{figure*}[t]
    \centering
    \includegraphics[width=\linewidth]{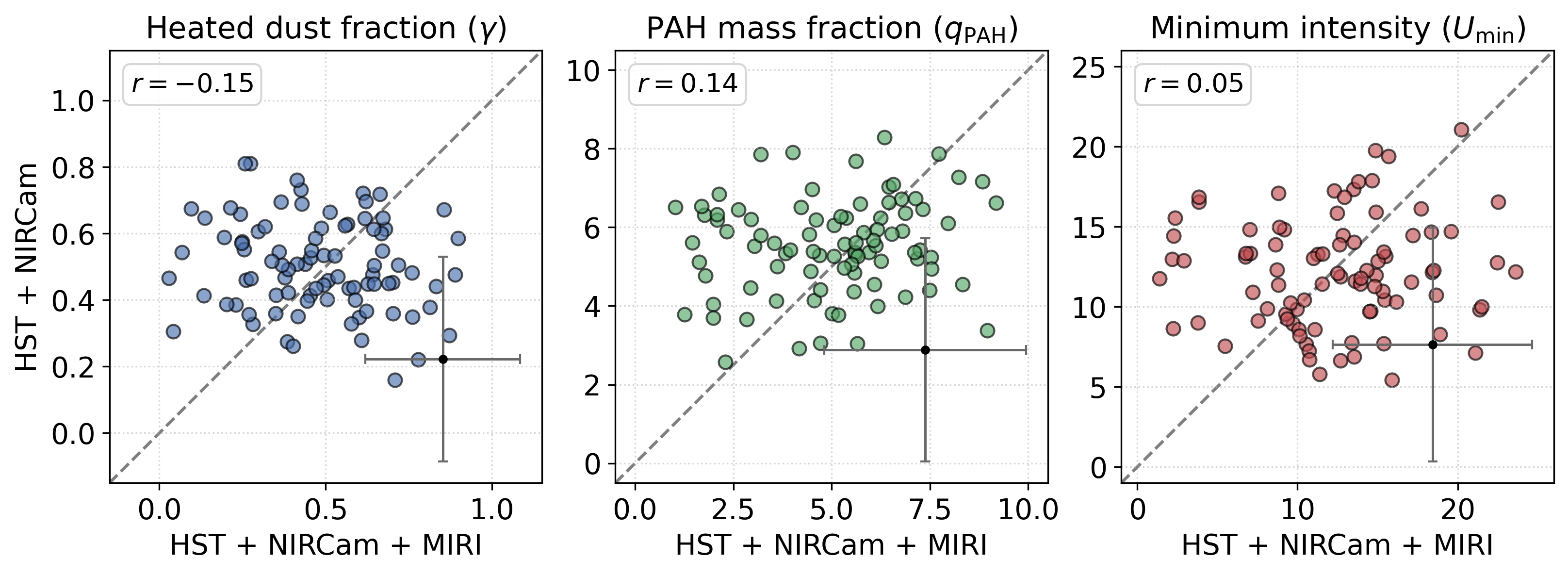}
    \caption{Comparison of the median dust emission parameters of the \citet{draine2007dust} dust model when fitting photometry from HST and NIRCam with and without MIRI data. The Pearson correlation coefficients are shown in the top left corner of each panel, while the median uncertainties are displayed in the bottom right corner.}
    \label{fig:duste_params_miri}
\end{figure*}

In principle, \textsc{Prospector} can lower mid-infrared fluxes in two ways: (i) by reshaping the dust emission at fixed total energy and shifting emission towards the far-infrared (e.g. by varying the starlight intensities, PAH fractions, dust mass and temperature), or (ii) reducing the total amount of absorbed starlight that is re-radiated by dust. We first examine the median dust emission parameters of the \citet{draine2007dust} dust model obtained from the photometric fits with and without MIRI and show the results in Fig.~\ref{fig:duste_params_miri}. All of the best-fit parameters appear to be randomly distributed, filling the entire prior space and showing no correlation between the different configurations. This behaviour could suggest that MIRI photometry effectively reduces the prior volume and converges on galaxy-specific dust emission properties. Alternatively, it may indicate that broadband mid-infrared photometry alone is unable to decouple the underlying dust emission parameters at cosmic noon.

To distinguish between these two scenarios, we performed parameter recovery tests using mock galaxy SEDs with known input dust properties generated via \textsc{Bagpipes} \citep{carnall2018bagpipes} and re-fitted independently with both \textsc{Prospector} and \textsc{Bagpipes}. The full methodology is detailed in Appendix~\ref{app:mock_fits}. Crucially, we assumed an idealised observational setup in which we have access to all nine MIRI imaging bands (F560W through F2550W), alongside the HST and NIRCam filters discussed in this work. We simulated realistic observational conditions by perturbing the synthetic photometry, assuming 10\% Gaussian flux errors. We then refitted the perturbed photometry for two mock galaxies at $z = 1.0$ and $z = 2.5$, respectively. At $z=1.0$, the MIRI bands cover the prominent 6.2, 7.7, and $8.6\,\mu\mathrm{m}$ PAH features, allowing both fitting codes to recover the true PAH mass fraction ($q_{\mathrm{PAH}}$) within $1\sigma$. However, at $z=2.5$, many of the PAH features shift out of the most sensitive MIRI bands, and the rest-frame mid-infrared continuum is sampled by only a handful of filters on the Wien tail of the dust emission. Under these conditions, both codes fail to accurately recover the input value of $q_{\mathrm{PAH}}$, although we find that \textsc{Bagpipes} yields more accurate estimates. We find that the remaining dust emission parameters are poorly constrained in both fits, indicating that the large scatter observed in Fig.~\ref{fig:duste_params_miri} is largely driven by a data limitation rather than improved estimates of the dust properties at cosmic noon. 

\subsection{The Impact on Total Infrared Luminosity $L_{\mathrm{IR}}$}
\label{ssc:lir}

Lastly, we investigate the impact of including MIRI photometry on the total infrared luminosities ($L_{\mathrm{IR}}$) inferred from the SED models. This will allow us to test whether the lower mid-infrared fluxes reflect a spectral reshaping of the dust emission templates (e.g. by varying the PAH fraction $q_{\mathrm{PAH}}$ or starlight intensity $U_{\min}$ at fixed total energy) or a genuine reduction in the total absorbed-and-re-emitted energy budget ($L_{\mathrm{abs}}$). To this end, we compare two distinct photometric fitting configurations: a fiducial fit using only HST and NIRCam photometry that we use to calculate $L_{\mathrm{IR}}^{\mathrm{fiducial}}$, and a fit including the MIRI photometry yielding $L_{\mathrm{IR}}^{\mathrm{MIRI}}$.

To compute the infrared luminosities, we first draw 100 posterior-weighted SED models from each fiducial and MIRI fit, respectively. For each SED sample we calculate the total infrared luminosity
\begin{equation}
    L_{IR} = L_{8-1000\,\upmu \mathrm{m}} = \int_{8\,\upmu \mathrm{m}}^{1000\,\upmu \mathrm{m}} L_\lambda\,d\lambda\,,
\end{equation}
following the prescription of \citet{sanders1996luminous}. Lastly, for each galaxy we extract the median luminosity from all samples, together with the 16th and 84th percentile to obtain $L_{\mathrm{IR}}^{\mathrm{MIRI}}$ and $L_{\mathrm{IR}}^{\mathrm{fiducial}}$, respectively. 

\begin{figure}[t]
    \centering
    \includegraphics[width=\hsize]{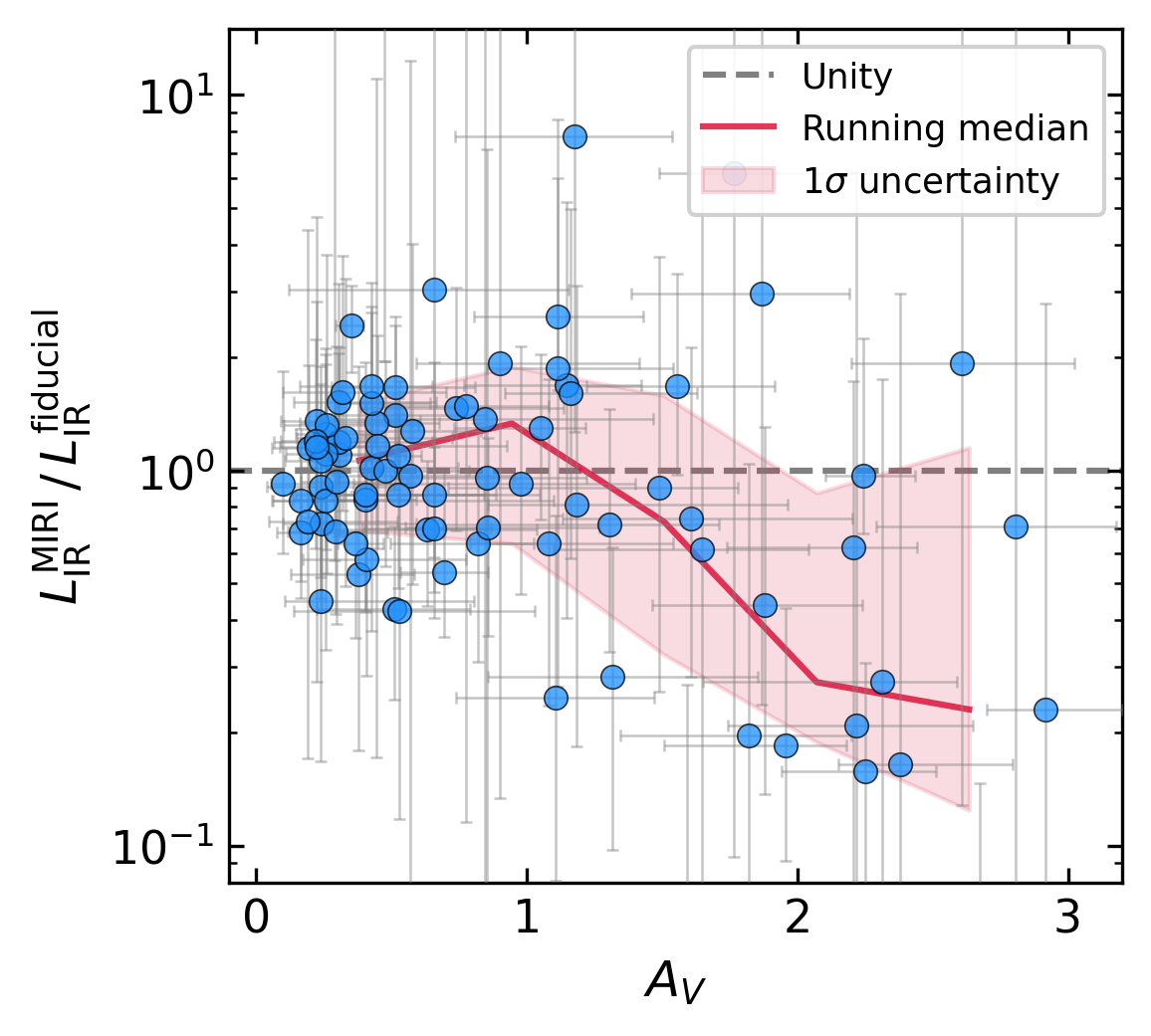}
    \caption{Ratios of total infrared luminosities $L_{8-1000\,\mu m}$ inferred from photometric fits with and without MIRI. Inferred dust attenuation ($A_V$) is displayed on the x-axis. The running median and the 1$\sigma$ uncertainty are indicated in red.}
    \label{fig:lir_ratios}
\end{figure}

Figure~\ref{fig:lir_ratios} displays the ratio of total infrared luminosities ($L_{\mathrm{IR}}^{\mathrm{MIRI}} / L_{\mathrm{IR}}^{\mathrm{fiducial}}$) as a function of the diffuse dust attenuation ($A_V$). While low-attenuation galaxies ($A_V \lesssim 1.0$) cluster near unity with moderate scatter, we observe a pronounced, systematic decline towards higher dust columns. For heavily attenuated systems ($A_V \gtrsim 1.0$), the running median drops steeply to~$\sim0.25$, corresponding to a systematic reduction of $\approx 0.6$ dex. 
This confirms that fitting the available MIRI photometry effectively reduces the amount of absorbed and re-emitted starlight rather than reshaping the dust emission template to match the observations.

This discrepancy yields important consequences for derived galaxy scaling relations at cosmic noon. In particular, star formation rates inferred from energy-balance SED modelling ($\mathrm{SFR} \approx \mathrm{SFR}_{\mathrm{UV}} + \mathrm{SFR}_{\mathrm{IR}}$) in dusty systems could be overestimated without direct constraints on the mid-infrared emission. Furthermore, because the rest-frame UV luminosity ($L_{\mathrm{UV}}$) and spectral slope ($\beta$) are identical in both fitting runs, this systematic reduction in $L_{\mathrm{IR}}$ can be directly translated into a decrease of up to $\sim 0.6\,\mathrm{dex}$ in the infrared excess ($\mathrm{IRX} \equiv \log_{10}(L_{\mathrm{IR}} / L_{\mathrm{UV}})$). Consequently, dusty galaxies at cosmic noon can occupy different regimes on the $\mathrm{IRX}\text{--}\beta$ plane, depending on the fitting configuration. We find that including mid-infrared observations in the fitting process effectively pushes galaxies below the Calzetti-like attenuation law inferred from optical-only photometric fits, consistent with steeper, SMC-like dust curves (\citealp[e.g.][]{gordon2003smc}) or complex star--dust geometries. While older stellar populations and higher metallicities can also produce intrinsically redder UV slopes at lower $\mathrm{IRX}$, breaking these degeneracies requires fitting rest-frame optical spectroscopy alongside panchromatic photometry to isolate the true attenuation law.


\section{Summary and Conclusions}   \label{sc:discussion}

We present the Blue Jay MIRI photometric catalogue\footnote{\href{https://doi.org/10.5281/zenodo.24204865}{https://doi.org/10.5281/zenodo.24204865}} discussed in Appendix~\ref{app:phot_table_description}, containing public JWST/MIRI observations for 111 galaxies between $1.7<z<3.5$ in the COSMOS field. We establish a robust, reproducible photometric workflow for JWST/MIRI observations and use the measured photometry to test the assumption of dust energy balance for a subsample of 89 star-forming galaxies, as implemented in many modern spectral energy distribution (SED) fitting codes. Here we provide a brief summary of this work and discuss the results below:

\begin{itemize}
    \item We processed publicly available image mosaics from the PRIMER, COSMOS-Web, and COSMOS-3D surveys across four MIRI bands (F770W, F1000W, F1800W, and F2100W). To ensure consistency with the existing Blue Jay catalogue, we performed aperture photometry using rescaled NIRCam apertures to reduce the impact of PSF corrections on the measured fluxes. We applied custom astrometric corrections and employed an iterative local background modelling and source masking routine to carefully mitigate the effects of large-scale background gradients, detector artefacts, and bad pixels. The full photometric pipeline is publicly available in a custom-made Python package \texttt{miriutils}\footnote{\href{https://doi.org/10.5281/zenodo.24204809}{https://doi.org/10.5281/zenodo.24204809}}. We release the Blue Jay MIRI Photometric Catalogue, providing aperture-corrected fluxes, uncertainties, aperture parameters, and quality-control flags.

    \item For a selected sample of 89 star-forming galaxies we compare the measured MIRI photometry to predictions from \textsc{Prospector} best-fit model spectra obtained by fitting HST and NIRCam photometry. We find that mid-infrared fluxes are overestimated by about 10\% in F770W up to a factor of~$\sim3$ in F2100W.
    
    \item We refit the HST and NIRCam photometry with NIRSpec spectroscopy, which reduces overpredictions by $\sim50\%$ in the reddest MIRI bands. In a last fitting run using HST, NIRCam and MIRI photometry, we find that the infrared luminosities of dusty star-forming galaxies ($A_V\gtrsim1$) are lowered by up to $\sim0.6$ dex compared to the optical-only photometric fits, consistent with complex star--dust geometries or a steeper, SMC-like attenuation curve.
\end{itemize}

Our analysis demonstrates that SED models relying on dust energy balance reliably predict mid-infrared emission from rest-frame UV--optical photometry in relatively unobscured systems ($A_V \lesssim 1$) and at shorter wavelengths ($\lambda_{\mathrm{rest}} \lesssim 3\,\mu\mathrm{m}$), where the stellar continuum still contributes significantly. At longer wavelengths, where the model relies increasingly on the amount of absorbed starlight ($L_{\mathrm{abs}}$) to extrapolate the infrared continuum, overpredictions become significant and correlate strongly with inferred dust attenuation. We find that incorporating medium-resolution spectroscopy from JWST/NIRSpec breaks fundamental degeneracies between stellar age, metallicity, and dust attenuation along the line of sight. Together with tighter constraints on the total absorbed energy, the resulting best-fit SEDs better reproduce the observed MIRI fluxes and reduce the overpredictions by a factor of~$\sim2$ compared to fits using only HST and NIRCam photometry.

To understand the physical origin of these flux overpredictions, we consider two ways in which \textsc{Prospector} can lower mid-infrared fluxes: (i) by reducing the total amount of absorbed starlight that is re-radiated by dust, or (ii) by reshaping the dust emission spectrum at fixed total energy and shifting emission towards the far-infrared (e.g. by lowering radiation field intensities, PAH fractions, or dust temperatures).

Addressing the first scenario, we examine possible line-of-sight effects coupled to a complex star--dust geometry, by which viewing a galaxy edge-on can increase the inferred visual attenuation compared to the galaxy-averaged diffuse optical depth. The opposite is true if a galaxy is observed face-on, allowing a relatively unobscured view into the centre. Because dust emission is isotropic, extrapolating the global infrared emission from a single sightline based on dust energy balance can lead to systematic over- or underpredictions of the inferred energy output. \citet{maheson2025bigdusty} investigated the effect of disc inclination on inferred dust attenuation for a selected subsample of the Blue Jay survey and found a clear correlation of higher $A_V$ with larger stellar mass, while axis ratios appear to play a negligible role. Additionally, we find no evidence for a MIRI-band excess attributable to hidden dust clumps within the sensitivity and wavelength coverage of our MIRI data. Such clumps lying outside our direct line of sight do not contribute to the inferred dust attenuation, yet still radiate in the infrared. This would cause UV--optical fits to underpredict the observed MIRI fluxes, contrary to our findings.

Secondly, we investigate the possibility of an accurate absorbed energy budget whose emission is redistributed to longer wavelengths. In heavily obscured star-forming galaxies, high dust optical depths can induce strong self-shielding, whereby dense birth-cloud screens absorb energetic UV photons before they reach the diffuse ISM. This lowers the equilibrium dust temperature and shifts a larger fraction of the re-emitted energy toward the far-infrared peak rather than the warmer mid-infrared continuum while maintaining the same $L_{\mathrm{IR}}$. However, we find that fully photometric fits incorporating our MIRI measurements lower the inferred $L_{\mathrm{IR}}$ by up to $\sim0.6\,\mathrm{dex}$ in dusty systems ($A_V \gtrsim 1$) relative to optical-only fits. 

Because self-shielding is expected to preserve the total infrared luminosity, we deem this explanation less likely and instead conclude that the bulk of the discrepancy stems from a genuine overestimation of the total absorbed starlight due to particularly dusty sightlines through complex, clumpy interstellar media. Observing each galaxy along a single line of sight fails to capture its full three-dimensional star--dust geometry. Since dust energy balance assumes that this directional attenuation is representative of the entire galaxy, extrapolating the global infrared emission from a single heavily obscured sightline can systematically inflate the inferred absorbed energy budget.

Furthermore, the observed reduction in $L_{\mathrm{IR}}$ implies a downward shift on the IRX--$\beta$ plane, consistent with steeper, SMC-like attenuation curves or complex star--dust geometries. We note that while the \citet{draine2007dust} dust model successfully reproduces the observed mid-infrared fluxes, the individual dust emission parameters ($\gamma$, $q_{\mathrm{PAH}}$, and $U_{\min}$) remain poorly constrained, indicating that the models may be too flexible. We attribute the large scatter to data limitations rather than genuine physical diversity, as confirmed by mock SED fits discussed in Appendix~\ref{app:mock_fits}. At $z\gtrsim2$ MIRI broadband photometry misses key portions of the PAH complexes, as well as the entirety of the rest-frame far-infrared dust peak at $\lambda_{\mathrm{rest}}\sim100\upmu$m. We conclude that MIRI photometry acts as an important bolometric anchor to directly constrain the IR emission rather than providing a reliable diagnostic of dust grain properties at cosmic noon.

Overcoming the remaining degeneracies between PAH fractions, radiation field intensities, and dust temperatures at cosmic noon will require panchromatic spectroscopic and photometric coverage beyond broadband imaging with MIRI. A full characterisation of the PAH and dust properties at high redshift will require a community effort to study large galaxy samples using low- to medium-resolution mid-infrared spectroscopy (e.g. with JWST/MIRI MRS), as proposed by the Genesis initiative \citep{genesis2026multicycle}, to establish stronger empirical priors that can be applied beyond the local Universe. Additionally, combining MIRI photometry with sub-mm observations from facilities such as ALMA and NOEMA will be vital for constraining the shape of the dust continuum, while also providing better estimates of dust temperatures. Finally, forthcoming missions such as the PRobe far-Infrared Mission for Astrophysics \citep[PRIMA;][]{moullet2023prima, glenn2025prima} will allow us to bridge the gap between the warm dust seen by JWST and the cold dust continuum traced at sub-mm wavelengths. Until such multi-wavelength datasets are routine, incorporating direct mid-infrared photometry or deep rest-optical spectroscopy remains essential to quantify dust-obscured star formation and model the infrared emission in galaxies at cosmic noon.

\begin{acknowledgements}
      B.P.C. acknowledges support from the EPEC Doctorate Scholarship. 
      S.B., L.B., and A.H.K. are supported by ERC Starting Grant 101076080 "Red Cardinal". MP acknowledges support from the Kavli Institute for Cosmology, Cambridge and the Kavli Foundation, and King's College, Cambridge. VMB acknowledges support from JWST GO-01810. This work is based on observations made with the NASA/ESA/CSA James Webb Space Telescope. The data were obtained from the Mikulski Archive for Space Telescopes at the Space Telescope Science Institute, which is operated by the Association of Universities for Research in Astronomy, Inc., under NASA contract NAS 5-03127 for JWST. These observations are associated with programmes \#1837, \#1727, and \#5893. The Blue Jay data, including ancillary \emph{HST} and \emph{JWST} photometry, are publicly available; see \citet{belli2026bluejaysurveydeep} for details of the data release. Part of this work was supported by \emph{ESO}, project number Ts~17/2--1.
\end{acknowledgements}

\bibliographystyle{aa}
\bibliography{references}

%


\begin{appendix}




\onecolumn

\section{Astrometric Offsets}   \label{app:astrometry}
\renewcommand{\arraystretch}{1.5}
\begin{table*}[h]
    \centering
    \begin{tabular}{l l r r r r r r r}
        \hline\hline
        Survey & Filter & $N$ 
        & $\Delta$RA$_{\rm med}$
        & MAD$_{\rm RA}$
        & $\Delta$Dec$_{\rm med}$
        & MAD$_{\rm Dec}$
        & $|\Delta|_{\rm med}$\\
        \hline
        PRIMER-1   & F770W  &  8  & -0.168 & 0.032 &  0.046 & 0.028 & 0.174 \\
        PRIMER-1   & F1800W &  4  & -0.136 & 0.045 &  0.086 & 0.087 & 0.161 \\
        \hline
        PRIMER-2   & F770W  & 44  &  0.205 & 0.020 & -0.008 & 0.038 & 0.205 \\
        PRIMER-2   & F1800W & 18  &  0.168 & 0.027 &  0.023 & 0.061 & 0.170 \\
        \hline
        COSMOS-Web-1 & F770W  & 32 & -0.161 & 0.023 &  0.022 & 0.019 & 0.163 \\
        \hline
        COSMOS-Web-2 & F770W  & 12 &  0.223 & 0.023 & -0.006 & 0.042 & 0.223 \\
        \hline
        COSMOS-3D-1  & F1000W & 13 & -0.152 & 0.024 &  0.028 & 0.037 & 0.154 \\
        COSMOS-3D-1  & F2100W &  9 & -0.084 & 0.068 &  0.060 & 0.062 & 0.104 \\
        \hline
        COSMOS-3D-2  & F1000W & 18 &  0.245 & 0.044 & -0.010 & 0.017 & 0.245 \\
        COSMOS-3D-2  & F2100W &  9 &  0.214 & 0.056 & -0.026 & 0.035 & 0.216 \\
        \hline
    \end{tabular}
    \caption{Astrometric offsets between MIRI and NIRCam/F444W centroids for each observing run and filter. 
    Offsets are defined as $\Delta = \mathrm{MIRI} - \mathrm{NIRCam}$. The parameter \textit{N} indicates the number of sources used for astrometric comparison.
    The median absolute deviation (MAD) is reported as a robust estimate of the scatter. All values are given in arcseconds.}
    \label{tab:offsets}
\end{table*}

\section{Description of the Photometric Catalogue} \label{app:phot_table_description}

\begin{table*}[h]
    \centering
    \begin{tabular}{llll}
        \hline \hline
        Column Name & Unit & Data Type & Description \\
        \hline
        ID & -- & int64 & 3D-HST galaxy ID \\
        MIRI\_ap\_a & pix & float64 & Aperture semi-major axis in MIRI pixels \\
        MIRI\_ap\_b & pix & float64 & Aperture semi-minor axis in MIRI pixels \\
        MIRI\_ap\_npix & pix$^2$ & float64 & Number of pixels in the aperture \\
        FXXXX\_flux & Jy & float64 & Total flux (Aperture corrected) \\
        FXXXX\_flux\_err & Jy & float64 & 1-sigma flux uncertainty (Aperture corrected) \\
        FXXXX\_abmag & mag & float64 & AB magnitude \\
        FXXXX\_apflux & Jy & float64 & Aperture flux \\
        FXXXX\_apflux\_err & Jy & float64 & 1-sigma aperture flux uncertainty \\
        FXXXX\_apflux\_errnominal & Jy & float64 & Nominal error from weights ("\texttt{ERR}" extension) \\
        FXXXX\_apcorr & -- & float64 & Aperture correction factor \\
        FXXXX\_bkg & Jy & float64 & Background flux in the aperture \\
        FXXXX\_bkg\_err & Jy & float64 & MAD of the residuals after plane subtraction \\
        FXXXX\_qc\_flag & -- & int64 & Quality flag (0=CLEAN, 1=OVERSUB, 2=UNDERSUB, 3=ARTEFACT) \\
        FXXXX\_ap\_flag & -- & int64 & Aperture scaling flag (0=DEFAULT, 1=SMALL) \\
        FXXXX\_ap\_x & pix & float64 & Aperture X-coordinate in MIRI pixels \\
        FXXXX\_ap\_y & pix & float64 & Aperture Y-coordinate in MIRI pixels \\
        FXXXX\_ap\_theta & deg & float64 & Aperture rotation angle (counter-clockwise from positive x-axis) \\
        \hline
    \end{tabular}
    \caption{Description of columns in v2.0 of the Blue Jay MIRI photometric catalogue including the column name, unit (if applicable), data type and a brief description of its content. The prefix \texttt{FXXXX} corresponds to the MIRI filter used (e.g. F770W, F1000W, F1800W, or F2100W).}
    \label{tab:phot_table}
\end{table*}

\twocolumn

This appendix describes the structure and contents of the Blue Jay MIRI Photometric Catalogue v2.0, which provides mid-infrared photometry for the Blue Jay galaxy sample in the COSMOS field measured from JWST/MIRI observations obtained by the PRIMER, COSMOS-Web, and COSMOS-3D surveys. The catalogue is publicly available in FITS (binary) and ECSV (text) formats\footnote{\href{https://doi.org/10.5281/zenodo.24204865}{https://doi.org/10.5281/zenodo.24204865}}. All column metadata, including physical units, data types, and descriptions, are embedded in the file headers to ensure compatibility with TOPCAT and standard Python-based analysis workflows.

\subsection{Catalogue Structure}
Each row in the catalogue corresponds to a single galaxy in the Blue Jay sample. The table follows a wide-format structure in which global (band-independent) properties are stored once per source, whilst filter-dependent quantities are stored in dedicated columns for each MIRI band. Band-independent quantities include identifiers and aperture geometry parameters that remain fixed across all filters. Band-dependent quantities include aperture-corrected flux densities, associated uncertainties, aperture corrections, background statistics, aperture position (in pixels) and rotation angle (in the detector reference frame). These are stored in columns named according to the filter (e.g.\ \texttt{F770W\_flux} or \texttt{F770W\_flux\_err}). All flux densities are provided in units of Jy and undefined values are masked using Astropy's \texttt{MaskedColumn} functionality. A complete list of all catalogue columns and their descriptions is provided in Table~\ref{tab:phot_table}.

\subsection{Quality Flagging Based on Curve-of-Growth Analysis}
\label{app:qc_flags}

Quality flags are assigned to each source and filter based on the behaviour of the curve of growth (CoG). The CoG is obtained by measuring the cumulative flux within a sequence of increasing elliptical apertures centred on the source. For well-behaved background subtraction, the CoG is expected to increase monotonically at small radii and converge towards a stable plateau at large radii. We analyse the CoG profile to identify deviations from this expected behaviour that may indicate imperfections in the background modelling routine. The classification is based on the following diagnostics:

\begin{itemize}
    \item The sign and stability of the cumulative flux at large radii
    \item The fraction of aperture steps contributing positive or negative incremental flux
    \item The location of the CoG maximum relative to the photometry aperture
    \item The presence of significant dips or non-monotonic features after the peak flux is reached
\end{itemize}

Based on these criteria, one of four quality flags is assigned:

\begin{description}
    \item CLEAN (0): The CoG shows a physically plausible behaviour, reaching a stable plateau at large radii without significant deviations.

    \item OVERSUB (1): The cumulative flux decreases systematically at large radii and becomes significantly negative with respect to the photometric uncertainty. This behaviour is consistent with background over-subtraction.

    \item UNDERSUB (2): The CoG continues to rise without reaching a plateau, suggesting a positive contribution from the background due to background under-subtraction.

    \item ARTEFACT (3): The CoG displays a significant dip or non-monotonic feature after reaching its maximum, indicative of localised background structures or detector artefacts affecting the measurement. Such artefacts are often visible by eye and are most frequently observed in the F1800W and F2100W bands.
\end{description}

The quantitative thresholds used to define these regimes were calibrated empirically through visual inspection of representative CoG profiles across all filters. The flags are intended as diagnostic indicators and do not imply that the data are unusable. Any residual background structures or artefacts are fully captured by the empirical RMS, which in these cases will inflate the reported measurement uncertainties accordingly.

\onecolumn
\section{Observation and Detection Statistics}
\label{app:detection_stats}

\begin{figure*}[h]
    \centering
    \begin{subfigure}[t]{0.48\textwidth}
        \includegraphics[width=\linewidth]{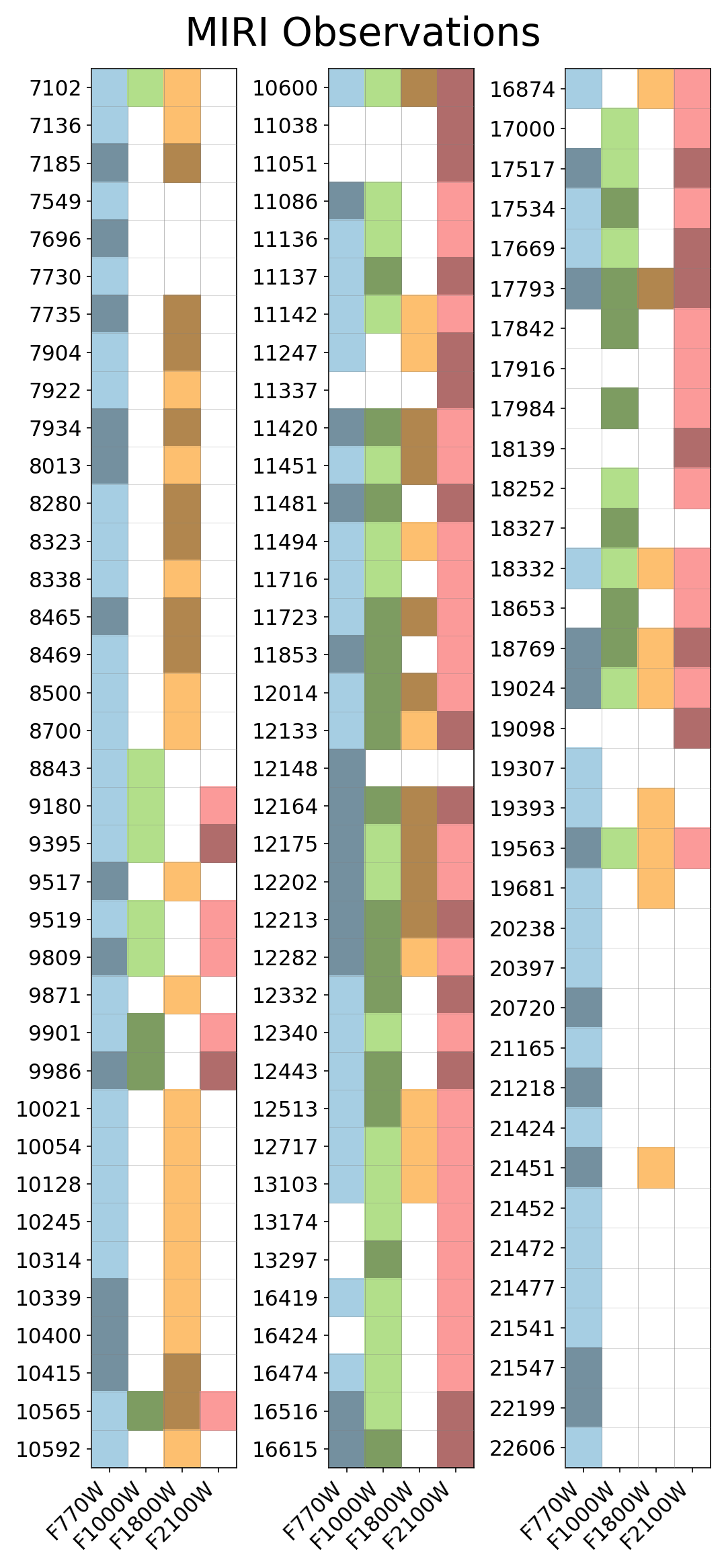}
        \caption{}
        \label{fig:miri_coverage}
    \end{subfigure}
    \hfill
    \begin{subfigure}[t]{0.48\textwidth}
        \includegraphics[width=\linewidth]{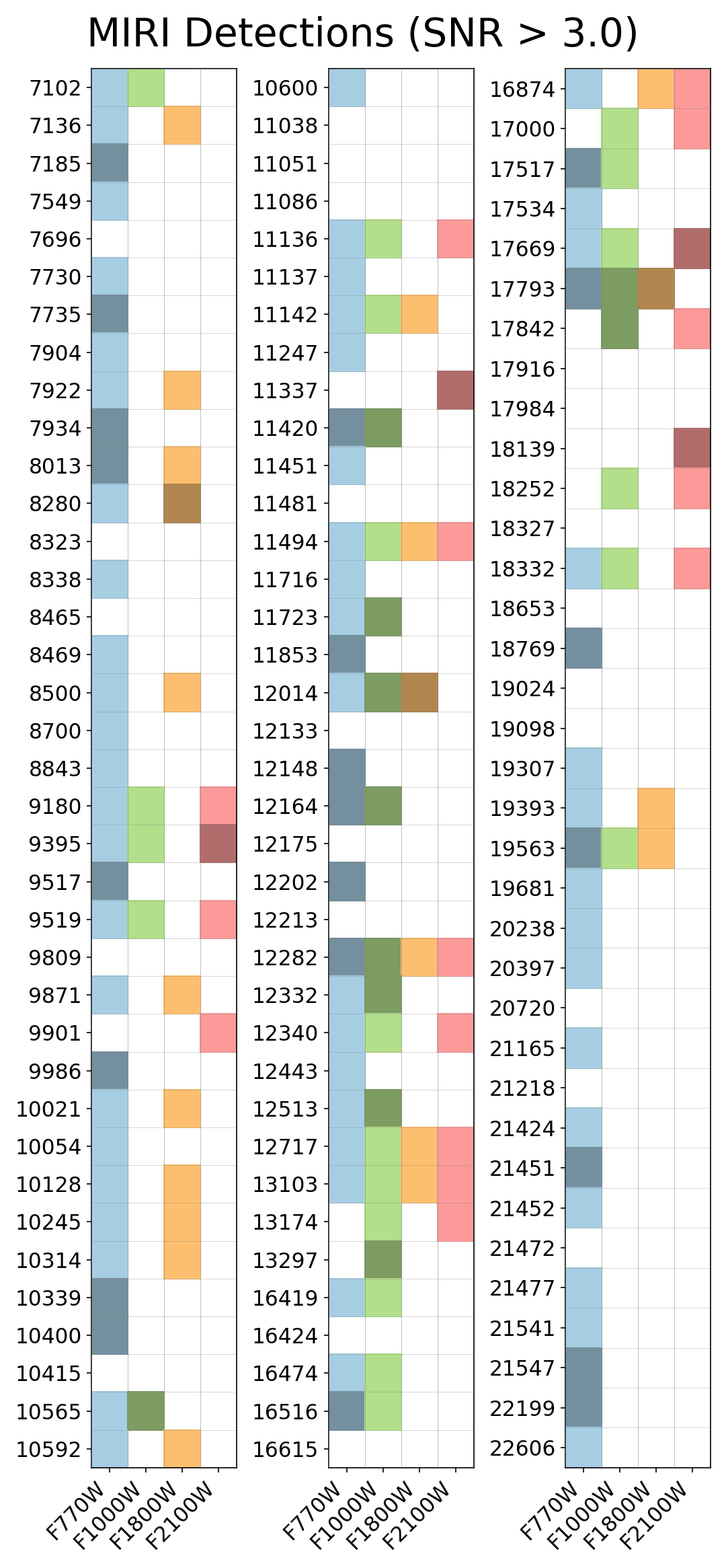}
        \caption{}
        \label{fig:miri_detections}
    \end{subfigure}
    \caption{Heat map showing (a) all MIRI filters available for each galaxy ID and (b) whether a galaxy is detected in a given band (S/N > 3). Dark patches indicate cutouts that triggered a flag (over-subtraction, under-subtraction or artefact).}
    \label{fig:miri_stats}
\end{figure*}

\twocolumn
\section{Constraining Dust Emission Parameters with MIRI}
\label{app:mock_fits}

To determine whether broadband JWST/MIRI photometry can uniquely constrain the dust grain population at high redshift, we performed a parameter recovery test using simulated galaxy photometry. We generated synthetic rest-frame UV-through-infrared spectra using the SED-fitting code \textsc{Bagpipes} \citep{carnall2018bagpipes}, adopting the \citet{charlot2000dust} dust absorption and the \citet{draine2007dust} dust emission models. Using the parameters shown in Table~\ref{tab:mock_parameters}, we generated the mock spectra of a galaxy that we placed at two representative redshifts: $z = 1.0$ and $z = 2.5$. In addition to the HST and NIRCam bands used in this work, we compute the mock photometry for an idealised configuration of all nine MIRI imaging filters (F560W through F2550W) and simulated observational errors by perturbing the photometry around $10\%$ of the model flux in each band.

\begin{table}[h!]
    \small
    \centering
    \renewcommand{\arraystretch}{1.2}
    \begin{tabular}{l l r}
        \hline\hline
        \noalign{\smallskip}
        Parameter & Description & Value \\
        \noalign{\smallskip}
        \hline
        \noalign{\smallskip}
        $\log_{10}(M_\star / \mathrm{M}_\odot)$   & Stellar mass                             & $11.00$ \\
        $\log_{10}(Z / \mathrm{Z}_\odot)$       & Stellar metallicity                      & $-0.40$ \\
        $A_V$                                   & Visual dust attenuation (mag)            & $3.00$  \\
        $n_{\mathrm{dust}}$                      & Dust attenuation power-law index         & $0.50$  \\
        $\gamma$                              & Fraction of dust in high-radiation field & $0.20$  \\
        $q_{\mathrm{PAH}}$                      & PAH dust mass fraction (\%)              & $1.50$  \\
        $U_{\min}$                              & Minimum radiation field intensity        & $7.50$  \\
        $\log_{10}(U)$                          & Ionisation parameter                     & $-3.00$ \\
        \noalign{\smallskip}
        \hline
    \end{tabular}
    \caption{Input physical parameters adopted for the synthetic galaxy SED.}
    \label{tab:mock_parameters}
\end{table}

\begin{figure}
    \centering
    \includegraphics[width=\linewidth]{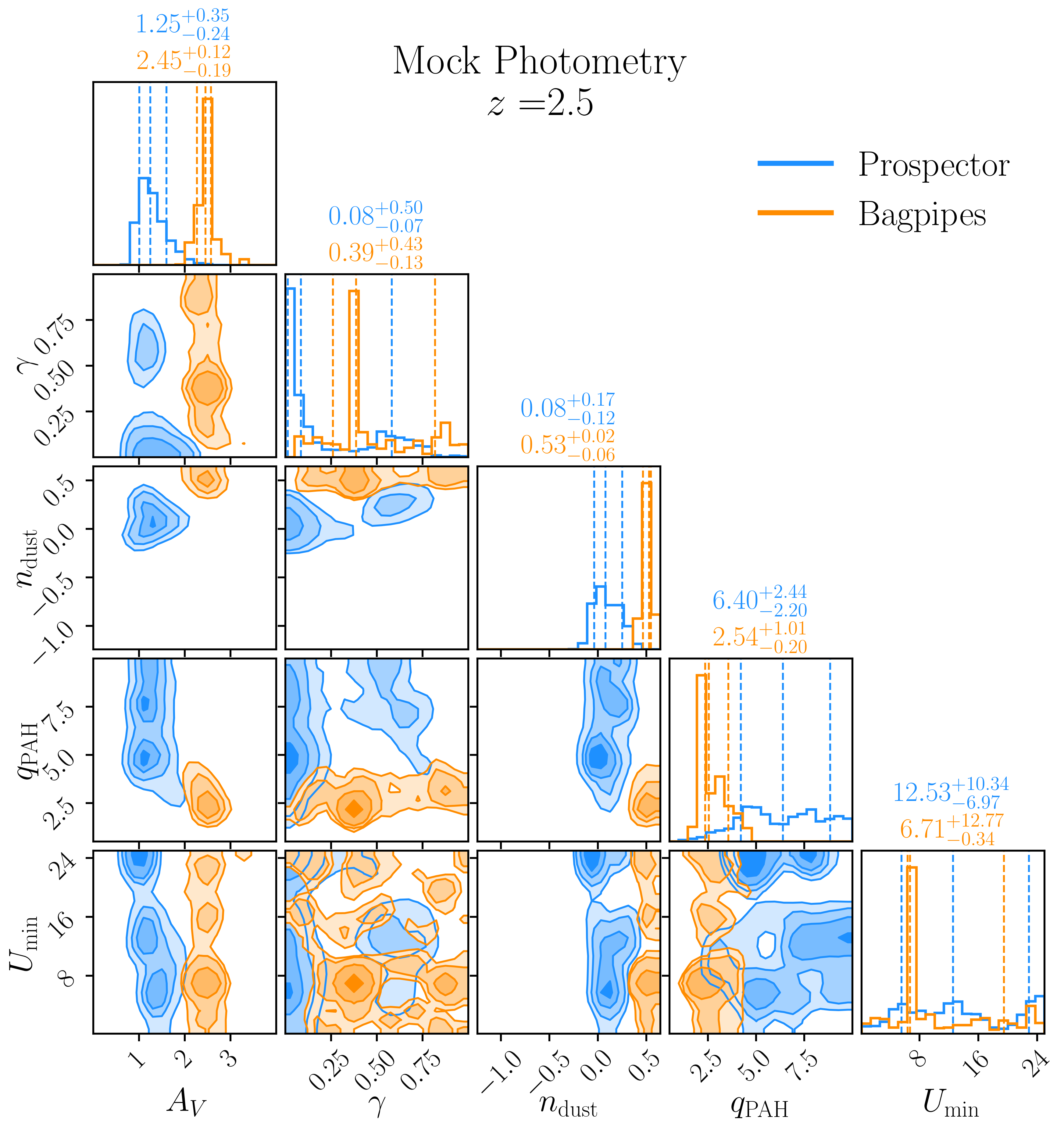}
    \caption{Posterior distribution of the dust parameters from \textsc{Prospector} (blue) and \textsc{Bagpipes} (orange) for a mock galaxy at $z=2.5$. The true values are listed in Table \ref{tab:mock_parameters}. Both codes fail to recover the input values, leaving the dust emission parameters largely unconstrained.}
    \label{fig:mock_z25}
\end{figure}

\begin{figure}
    \centering
    \includegraphics[width=\linewidth]{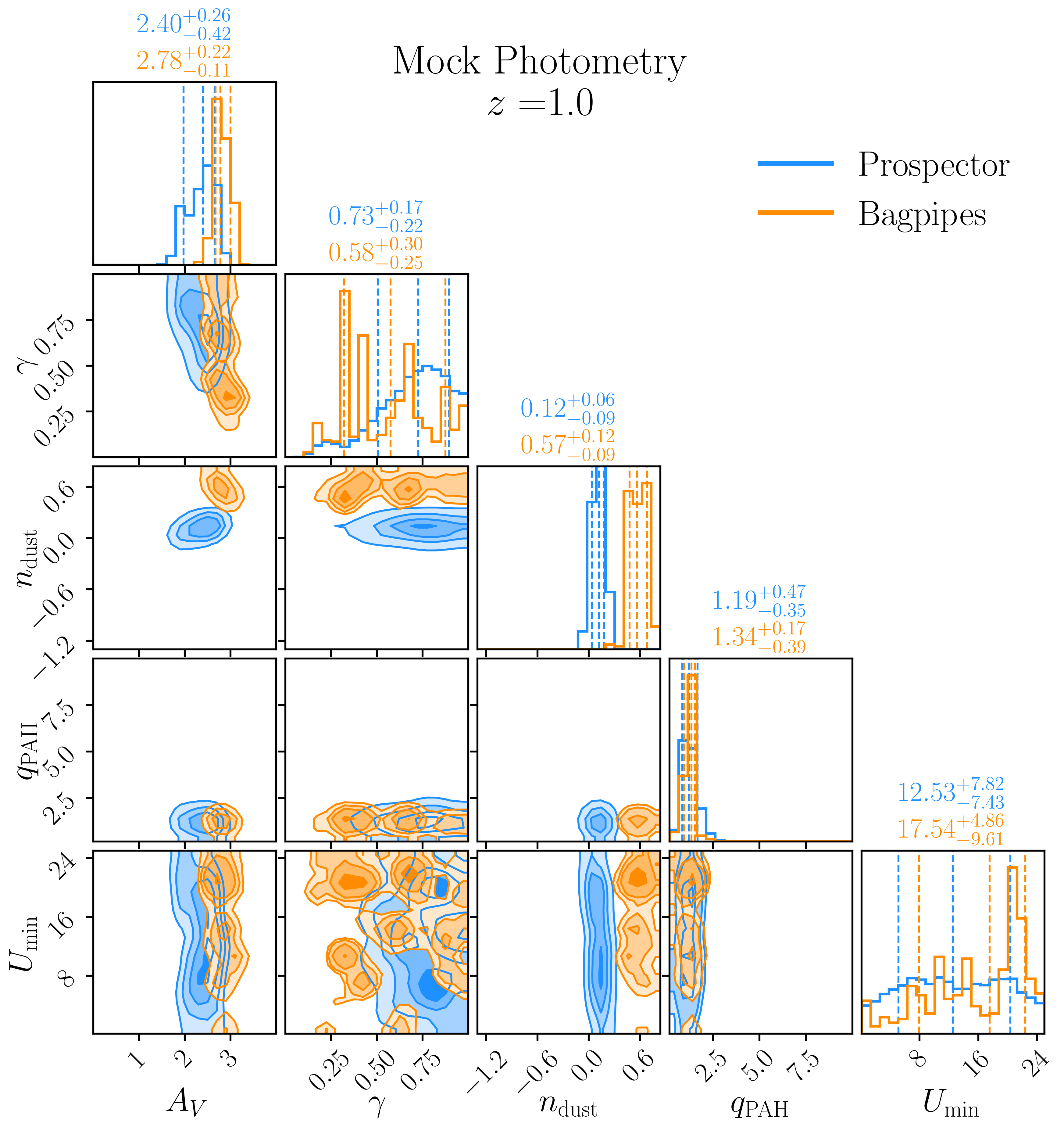}
    \caption{Posterior distribution of the dust parameters from \textsc{Prospector} (blue) and \textsc{Bagpipes} (orange) for a mock galaxy at $z=1.0$. The true values are listed in Table \ref{tab:mock_parameters}. Both codes show much narrower distributions in almost all parameters and $q_{\mathrm{PAH}}$ is recovered, while $\gamma$ and $U_{\mathrm{min}}$ remain unconstrained.}
    \label{fig:mock_z10}
\end{figure}

To exclude a possible dependence on the used SED fitting code, we fit the mock observations independently using both \textsc{Prospector} and \textsc{Bagpipes}, matching the SED setup described in Section \ref{ssc:sed_model_setup}. The resulting posterior parameter distributions are compared in Fig.~\ref{fig:mock_z25} ($z = 2.5$) and Fig.~\ref{fig:mock_z10} ($z = 1.0$). Parameters such as stellar mass and metallicity are reliably recovered by both codes, so they will not be discussed here.

At $z = 2.5$ the individual PAH features ($6.2\upmu$m, $7.7\upmu$m, and $8.6\upmu$m) are only partially sampled by MIRI, with many of the longer wavelength features shifting beyond its coverage. Under these conditions, the posterior probability distributions for $q_{PAH}$, the minimum starlight intensity $U_{\min}$, and the fraction of dust heated by high-intensity radiation $\gamma$ are poorly constrained, filling nearly the entire prior volume in both codes. At $z = 1.0$ (Fig.~\ref{fig:mock_z10}), the better coverage of the PAH complexes enables both codes to successfully recover the input dust mass fraction in PAHs ($q_{\mathrm{PAH}} = 1.19^{+0.47}_{-0.35}\%$ for \textsc{Prospector} and $1.34^{+0.17}_{-0.39}\%$ for \textsc{Bagpipes}, matching the true value of $1.50\%$). However, the remaining dust parameters remain loosely constrained, indicating that broad-band photometry alone is not sufficient to characterise the complex dust properties at high redshift.

The inability of both independent SED modelling codes to recover the input dust emission parameters at $z = 2.5$ (and largely also $z=1.0$) confirms that the large scatter seen in the updated dust emission parameters (Fig.~\ref{fig:duste_params_miri}) originates from a data limitation rather than an improved estimate of the true dust properties. 

\end{appendix}

\end{document}